\documentclass[aps,prb,groupedaddress,showpacs,floatfix,twocolumn]{revtex4}%
\usepackage{amsmath,amssymb}
\usepackage{graphicx}
\usepackage{dcolumn}
\usepackage{enumerate}
\usepackage{comment}
\usepackage{color}
\newcommand{\rmd}{\mathrm{d}}
\newcommand{\idMdB}{\rmd M{/}\rmd B}
\newcommand{\dMdB}{$\idMdB$}

\newcommand{\irJJ}{J^{(2)}{/}J^{(1)}}
\newcommand{\rJJ}{$\irJJ$}
\newcommand{\irJ}[1]{J_{#1}{/}J_1}
\newcommand{\rJ}[1]{$\irJ{#1}$}
\newcommand{\inmax}{n_{\max}}
\newcommand{\nmax}{$\inmax$}
\newcommand{\J}[1]{$J_{#1}$}
\newcommand{\JJ}[1]{$J^{({#1})}$}
\newcommand{\Spin}{$S{=}5{/}2$}
\newcommand{\muB}{\mu_\mathrm{B}}

\newcommand{\Fig}[1]{Fig.~\ref{Fig#1}}
\newcommand{\Figure}[1]{Figure~\ref{Fig#1}}
\newcommand{\Figs}[2]{Figs.~\ref{Fig#1} and \ref{Fig#2}}
\newcommand{\Figures}[2]{Figures~\ref{Fig#1} and \ref{Fig#2}}
\newcommand{\refone}{Bindi05prb}
\newcommand{\Refone}{Ref.~\onlinecite{\refone}}
\newcommand{\reftwo}{Bindi06eprint1}
\newcommand{\Reftwo}{Ref.~\onlinecite{\reftwo}}
\newcommand{\review}{Shapira02jap}
\newcommand{\Review}{Ref.~\onlinecite{\review}}
\newcommand{\refthree}{Gratens06prb}

\newcommand{\Sk}{\mathrm{S}}
\newcommand{\D}{\mathrm{D}}
\newcommand{\F}{\mathrm{F}}
\newcommand{\NTOT}{N^\mathrm{TOT}}
\newcommand{\NHF}{N^\mathrm{HF}}
\newcommand{\NLF}{N^\mathrm{LF}}
\newcommand{\skel}[1]{$(#1)_\Sk$}
\newcommand{\listpar}{\setlength{\topsep}{0pt}\setlength{\itemsep}{-
0.25\baselineskip}}
\begin{document}
\bibliographystyle{apsrev}
\title{Exchange-bond structure and magnetization-step spectra of a diluted
Heisenberg antiferromagnet on the square lattice:  Lopsided two-exchange  models}
\author{Yaacov Shapira}
\email{yshapira@granite.tufts.edu} \affiliation{Department of
Physics and Astronomy, Tufts University, Medford, MA 02155}
\author{Valdir Bindilatti}
\email{vbindilatti@if.usp.br} \affiliation{Instituto de
F\'{\i}sica, Universidade de S\~{a}o Paulo,\\ Caixa Postal 66.318,
05315--970 S\~{a}o Paulo-SP, Brazil}
\date{\today}
\begin{abstract}
This  paper on the theory of magnetization steps (MST's) from a strongly diluted
antiferromagnet on the square lattice follows two others on the same topic.
The preceding paper discussed Heisenberg models with two
antiferromagnetic exchange constants of arbitrary magnitudes.
The present paper specializes to ``lopsided'' models, in which the ratio
of the two exchange constants is so small that the MST spectrum has certain
characteristic features.
One important feature is a gap which divides the MST spectrum into a
low-field part and a high-field part.
The models that are considered are the lopsided  $J_1$-$J_2$ model and
the lopsided $J_1$-$J_3$ model, where $J_i$ is the exchange constant with
the $i$th neighbor.
The criteria that the ratios $J_2{/}J_1$ and $J_3{/}J_1$ must satisfy in order
for these
models to qualify as lopsided are obtained for magnetic ions with spin \Spin.
The bulk of the paper is devoted to a detailed exploration of the many
connections
between the exchange-bond structure of spin clusters and the MST spectra.
These connections are established by extensive numerical simulations of spectra,
supplemented by physical arguments.
The concepts of ``skeleton'' and ``decoration'' are central to the discussion of
the
exchange-bond structure of cluster types with one or more $J_1$ bonds.
A skeleton consists of the (strong) $J_1$ bonds and the spins attached to them.
Other spins and exchange bonds, if there are any, are in the decoration.
A skeleton is either ``simple'' or ``compound.'' A compound skeleton consists of
several ``fragments.''
Simple skeletons and fragments are classified by types, which fully specify the
$J_1$ bonds.
Clusters containing simple skeletons and fragments of one type produce a ``mono-
skeleton'' fine structure.
It consists of groups of very close lines in the high-field part of the
spectrum.
Spectral lines in the low-field part of the spectrum are related to the weak
exchange bonds.
Some of these lines arise from pure-$J_2$  (or pure-$J_3$) cluster types.
Others are related to $J_2$ (or $J_3$) bonds in decorations of mixed clusters.
The connections between these low-field lines and the exchange-bond structure
of the decorations are discussed in detail.
Two alternative methods of measuring the weak exchange constant are outlined.
One involves low-field lines from  $J_2$ (or $J_3$) pairs.
The other involves some high-intensity lines in the high-field fine structure.
The contribution of clusters of sizes $n_c{>}5$ to the magnetization is
approximated by the
corrective quintets (CQUIN's) method.
A detailed description of this method is given in an Appendix.
\end{abstract}
\maketitle

\section{\label{s:I-intro}INTRODUCTION}
This is the third theoretical paper on magnetization steps (MST's) from a
strongly diluted Heisenberg antiferromagnet on the square lattice.
The first paper\cite{\refone}  (also called I) was based on the nearest-neighbor
(NN) cluster model, or \J1 model.
The second paper\cite{\reftwo} (also called II) discussed ``generic'' and
``specific'' cluster models with two antiferromagnetic (AF) exchange constants.
The generic \JJ1-\JJ2 model placed no restriction on the ratio \rJJ\ between the
second-largest and largest exchange constants.
Similarly, there was no restriction on the ratio of the exchange constants in
the two specific models that were considered: 1) the \J1-\J2 and the \J1-\J3
models,
where $J_i$ is the exchange constant with the $i$th neighbor.

Numerical simulations of the magnetization curve, $M$ versus $B$, and of the
derivative curve, \dMdB\ versus $B$, were also carried out in II.
The pattern of the peaks in the derivative curve was defined as the ``MST
spectrum.''
For the \J1-\J2 model the simulated spectra indicated that when the ratio \rJ2
is ``very small'' the spectrum  has the following  features:
\begin{enumerate}\listpar
\item  The spectrum consists of a low-field part and a high-field part.
The two parts are separated by a ``gap'' in which there  are no discernable
spectral lines.
\item The magnetization curve  exhibits a plateau in the field region of the
gap. This plateau corresponds to the apparent  saturation in the \J1
model.\cite{\refone}
\item In the high-field part of the spectrum, each spectral line that exists in
the \J1 model may, and usually does, develop a fine structure (FS).
\item Separations of spectral lines in this FS are of order $\Delta b_2{\sim}1$,
where $b_2$ is the secondary reduced magnetic field, defined by Eq.~(3b) of~II.
The corresponding separations $\Delta B$ are of order  $|J_2|{/}g\muB$, where
$g$  is the g-factor and $\muB$  is the Bohr magneton.
\item  Typical line separations $\Delta b_2$ in the low-field part of the
spectrum are also of order   $1$.
\end{enumerate}
When the ratio \rJ2 in the \J1-\J2 model is sufficiently small that the
spectrum has the five features listed above, the model is called ``lopsided.''
The lopsided \J1-\J3 model is defined in a similar manner.
The relevant small ratio is then \rJ3, and line separations $\Delta B$ in
the high-field FS, and in low-fields, are controlled by \J3.

The present paper is devoted to the lopsided \J1-\J2 and \J1-\J3 models.
Among the topics that will be  discussed are:
\begin{enumerate}\listpar
\item  The numerical criteria that the ratios \rJ2 and \rJ3 must satisfy in
order for \J1-\J2  and \J1-\J3  models to qualify as lopsided;
\item The concepts of ``skeleton'' and ``decoration,'' for the
exchange-bond structure of clusters;
\item Connections between many features of the MST spectra and the relevant
skeletons and decorations;
\item Features of the spectrum that are most useful for  measuring the
second-largest exchange constant.
\end{enumerate}

As in I and II, the  present paper includes extensive numerical simulations of
MST spectra.
The simulations are only for magnetic ions with spin \Spin, and only for
zero temperature, $T{=}0$.
Some simulations at finite $T$ are included in the following paper as a part of
the data analysis.\cite{\refthree}

In calculations of the spectra (\dMdB\  vs $B$) the infinite sum in Eq.~(2)
of~II
is truncated at the maximum cluster size $\inmax{=}5$, and the remainder is
ignored.
In simulations of the magnetization $M$, on the other hand, the remainder is
approximated by  the corrective quintets (CQUIN's) method.
This method is described in Appendix~\ref{a:A}.

\section{\label{s:II}REQUIREMENTS FOR A MODEL TO QUALIFY AS LOPSIDED}
\subsection{\label{ss:IIA}The various criteria}
In order for a \JJ1-\JJ2 cluster model to qualify as ``lopsided,'' the ratio
\rJJ\ must
satisfy two numerical criteria: a  ``general criterion'' and   a ``gap
criterion.''
The general criterion is
\begin{equation}
\irJJ < 1/10.       \label{eq1}
\end{equation}
That is, the two exchange constants must differ by at least one order of magnitude.
The ``gap criterion'' requires the ratio \rJJ\ to be sufficiently small that
a gap, with no spectral lines, separates the low-field and  high-field
parts of the spectrum.

In contrast to the general criterion, the gap criterion depends on $S$,
on the maximum cluster size \nmax  at which the sum in Eq.~(2) of~II is
truncated,
and, in principle, also on the specific cluster model.
These dependencies cause the gap criterion to be more complicated than the
general criterion.
Both the general criterion and the gap criterion must be satisfied.
The more stringent of the two criteria is the ``governing criterion''
that a model is lopsided.

\subsection{\label{ss:IIB} Gap criterion for \Spin\ and $2{\le}\inmax{\le}5$}
The criteria for the existence of a gap in  the \J1-\J2 and \J1-\J3 models
were obtained only for  \Spin\  and only for $2{\le}\inmax{\le}5$.
It turned out that at least for these values of $S$ and \nmax, the
numerical criterion that  the ratio \rJ3 must satisfy in order for the \J1-\J3
model to be lopsided is the same as the criterion for the ratio \rJ2  in
the \J1-\J2 model.
Therefore, only the ratio \rJ2 in the \J1-\J2 model will be discussed in detail.

The first step was to obtain a necessary condition that the ratio \rJ2 must
fulfill  in order for
a gap to exist. This condition depends on \nmax.
Numerical simulations then showed that for $\inmax{=}2$ and $3$ the
necessary condition is also sufficient, so that it is the gap criterion.
For $\inmax{=} 4$ and $5$, however, the necessary condition is  not
quite sufficient. A minor adjustment was therefore  required to obtain the gap
criterion.

\subsubsection{\label{sss:IIB1} Necessary condition for a gap}
Cluster types that give rise to MST's in the \J1-\J2 model belong to three
categories:  pure-\J1, pure-\J2, and  ``mixed.'' Only the last category involves
both $J$'s.
The gap is in the total spectrum, from all cluster types of all categories.
The necessary condition for a gap was obtained by considering only the spectrum
from the pure cluster types, i.e., mixed cluster types were excluded.

The pure-\J2 cluster types produce spectral lines only in the low-field part of
the spectrum.
On the other hand, the pure-\J1 cluster types produce lines in the
high-field part of the spectrum.
A necessary condition for a gap is that the highest $B$ for any MST  from the
pure-\J2 cluster
types is below the lowest $B$  for any MST  from the pure-\J1 cluster types.
The pure-\J2 cluster types are all the cluster types of the  (parent) \J2 model.
The pure-\J1 cluster types are a known subset of the cluster types of the
(parent) \J1 model.
The two parent models are isomorphic (see Appendix A of~II).
For any choice of \nmax, only lines from cluster types with sizes
$n_c{\leq}\inmax$ are considered.

For \Spin, the primary reduced fields $b_1$ at the MST's from each cluster
type of  the \J1 model are given in Table II of~\Refone.
The secondary reduced fields $b_2$ at the MST's from the isomorphic cluster
types in the \J2 model are the same.
For cluster sizes not exceeding $\inmax{=} 2, 3, 4,$ and $5$, the highest value
of $b_2$ is $b_{2,\max}{=}10, 15, 20,$ and $25$, respectively.
The lowest $b_1$ is $b_{1,\min}{=}2$ when $\inmax{=} 2$ or  $3$,
and $0.950$ when $\inmax {=} 4$ or $5$.

Expressing  $b_{2,\max}$  and  $b_{1,\min}$  in terms of  $B$, the necessary
condition
for a gap is  $\irJ2{<}(b_{1,\min}{/} b_{2,\max})$.
Thus, the necessary condition for $\inmax{=} 2, 3, 4, 5$ is
$\irJ2{<}0.200, 0.133, 0.0475, 0.0380$, respectively.
\subsubsection{\label{sss:IIB2} The gap criterion}
The sufficient condition for a gap was obtained from numerical simulations
that also included the mixed cluster types.
For  $\inmax{=} 2$ and $3$, the necessary condition  was found to be sufficient.
Therefore, the gap criterion for $\inmax{=} 2$ is $\irJ2{<} 0.200$.
For  $\inmax{=} 3$,  the gap criterion is $\irJ2{<} 0.133$.

When \nmax\  is either $4$ or $5$, the sufficient condition for a gap is
slightly more stringent than the necessary condition.
The reason is that among the lines in the high-field part of the spectrum, the
line with the lowest $b_1$ is from a mixed cluster, not from  a pure-\J1
cluster.
The relevant mixed-cluster type is 4-3 (4-1 in the \J1-\J3 model).
Including the lowest $b_1$ from this cluster type,  the gap criterion for
$\inmax{=} 4$  is  $\irJ2{<} 0.0450$.
For $\inmax{=}5$, the gap criterion is $\irJ2{<} 0.0364$.

\subsection{\label{ss:IIC} Governing criterion for a lopsided model
when \Spin\  and  $2{\le}\inmax{\le}5$}
The more stringent of the general criterion and the gap criterion is the
governing criterion.
For   $\inmax{=}2$ and $3$ the governing criterion is the general criterion,
$\irJ2{<} 0.10$.
For $\inmax{=} 4$ and $5$, on the other hand,  the gap criterion is the
governing criterion.
The governing criterion is  therefore $\irJ2{<} 0.0450$ for $\inmax{=}4$, and
$\irJ2{<} 0.0364$ for $\inmax{=}5$.

\subsection{\label{ss:IID} Selecting \nmax}
The selection of \nmax\  in the calculation of the spectrum depends on several
considerations. One is that \nmax\  cannot exceed the largest size of clusters
whose
Hamiltonians can be diagonalized with available  resources.
Sometimes, however, \nmax\  is chosen to be lower than the upper limit of
technical
capabilities, because the labor of calculating the spectrum is reduced
substantially.

Because spectral lines from cluster types with sizes $n_c{>}n_\mathrm{max}$ are
lost
in the simulation, a choice of a lower \nmax\  implies a greater loss of
spectral lines.
However, this greater loss is  not always significant. If the concentration $x$
of the
magnetic ions  is sufficiently low, only a negligible fraction of the total
number of
magnetic ions are in clusters with $n_c{>}\inmax$.
In the present  paper,  all the calculated spectra are based on the choice
$\inmax{=}5$.

\section{\label{s:III}SKELETONS AND DECORATIONS}
\subsection{\label{ss:IIIA}Introductory remarks}
The MST spectrum can always be obtained numerically using the procedures
described in  \Reftwo.
Quite often, however, the numerically-obtained spectrum does not immediately
reveal some
of the underlying physics.
A major goal of the present paper is to expose the connections between many
features of the MST spectrum and the exchange-bond structures of the relevant
clusters.
The discussion below is mainly for the lopsided \J1-\J2 model.
The physics of the lopsided \J1-\J3 model is very similar.

Three categories of cluster types give rise to MST's in the \J1-\J2 model:
pure-\J2 cluster types, pure-\J1 cluster types, and mixed cluster types.
The spectra from the pure cluster types have already been discussed in
Sec.~\ref{ss:IIB}.
Therefore, the mixed cluster types are the main (but not the only) focus of the
remaining discussion in this paper.
The majority of all spectral lines are, in fact, from  mixed cluster types.

The concepts of ``skeleton'' and ``decoration'' are central to the discussion of
the exchange-bond structure  of mixed cluster types.
They are not essential concepts for pure-\J1 cluster types.
However, the use of these concepts both for mixed and for pure-\J1 cluster
types has the advantage of  a more uniform presentation.
The common feature of mixed and pure-\J1 clusters is that each such cluster
contains at least one \J1 bond.
In the physical picture that will be adopted here, any cluster with at least one
\J1 bond consists of a skeleton and a decoration.

\begin{figure*}\includegraphics[scale=0.9]{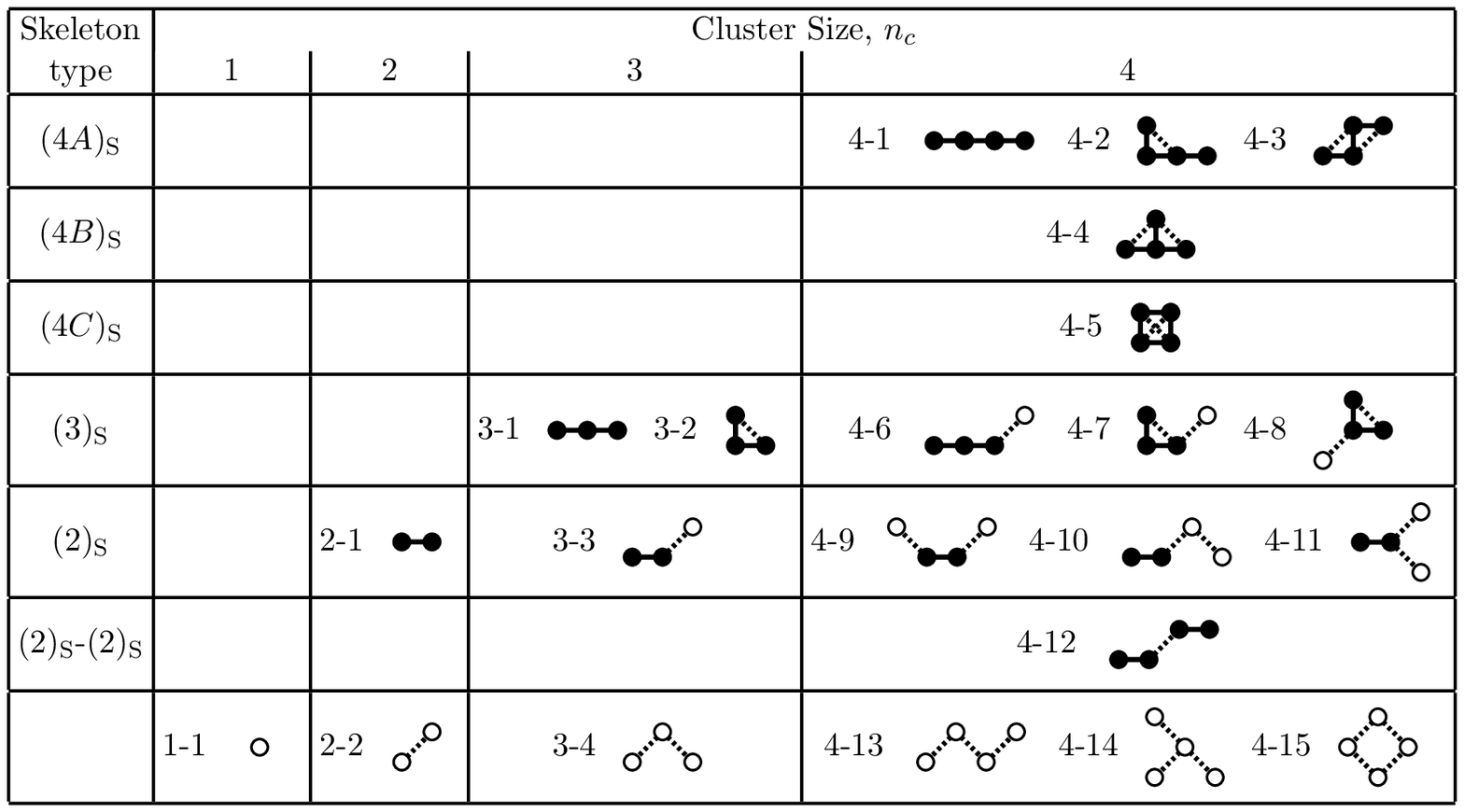}
\caption{\label{Fig1} Cluster types of the lopsided \J1-\J2 model, up to size
$n_c{=}4$.
This figure is similar to Fig.~1 of \Reftwo,  except that it emphasizes the
skeletons.
Skeleton-spins are shown as solid circles, whereas other spins are represented
by empty circles.
The solid lines  represent \J1 bonds. The \J2 bonds are shown as dotted lines.
The first column gives the skeleton type for all the skeletons in the same row.}
\end{figure*}

\begin{figure*}\includegraphics[scale=0.9]{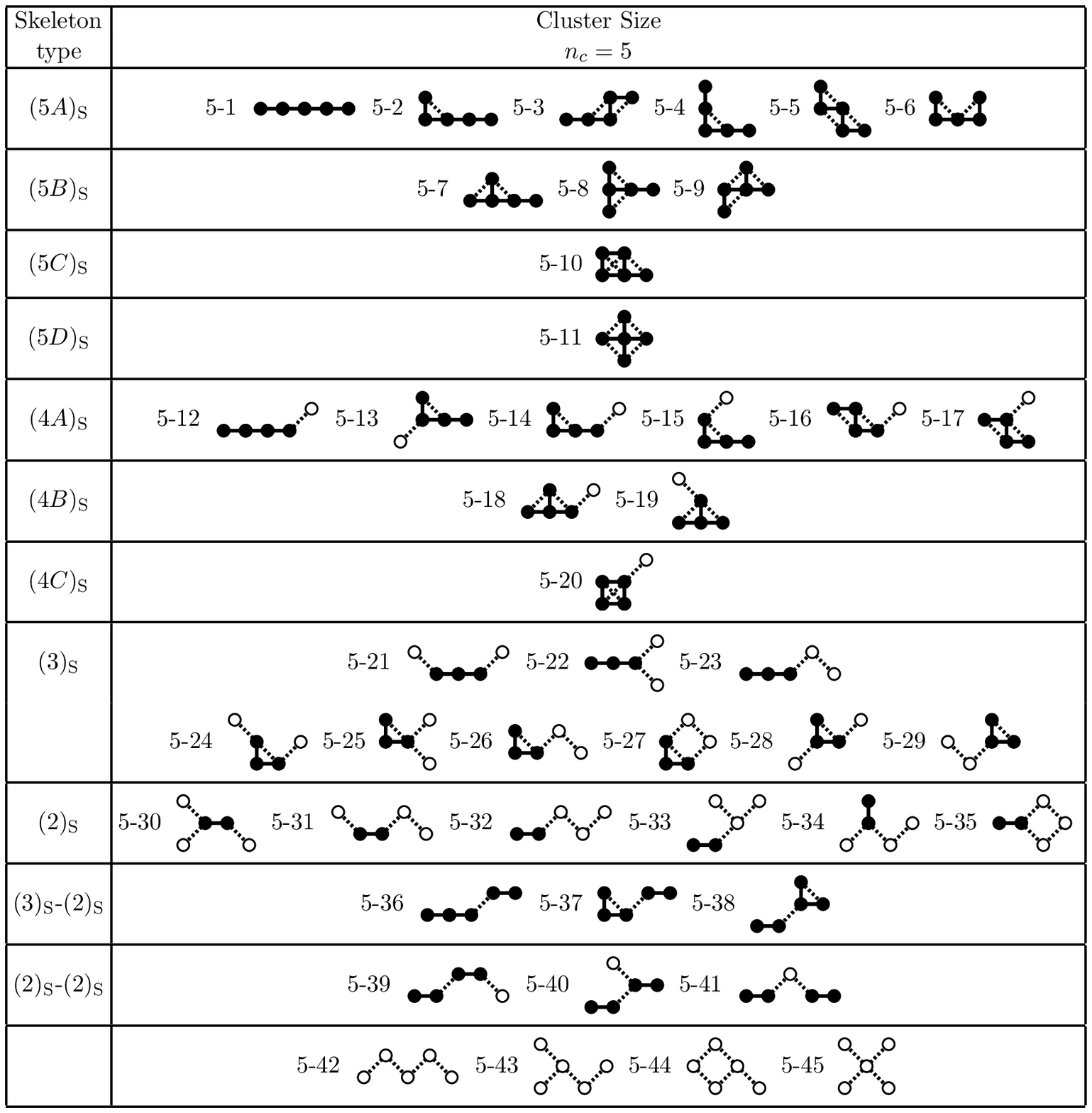}
\caption{\label{Fig2} The quintet types of  the lopsided \J1-\J2 model.
This figure is similar to Fig.~2 of \Reftwo\  except that the skeletons are
emphasized.
The first column gives the skeleton type.}
\end{figure*}

\begin{figure*}\includegraphics[scale=0.9]{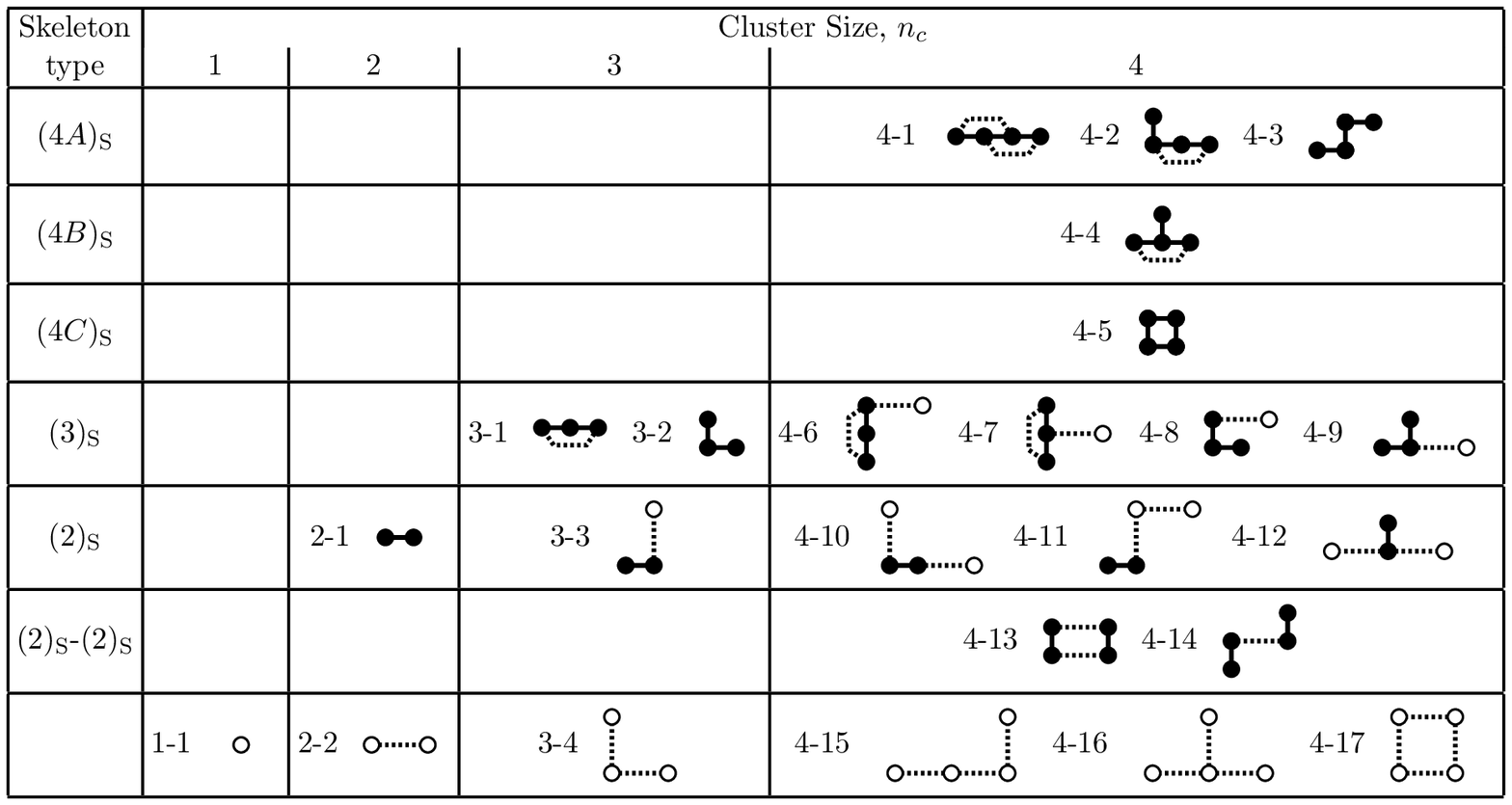}
\caption{\label{Fig3} Cluster types of the lopsided \J1-\J3 model, up to size
$n_c{=}4$.
The skeletons are emphasized by  using solid circles for skeleton-spins and
solid lines
for \J1 bonds. Dotted lines  represent \J3 bonds. The skeleton type is given in
the first column.}
\end{figure*}

\begin{figure*}\includegraphics[scale=0.9]{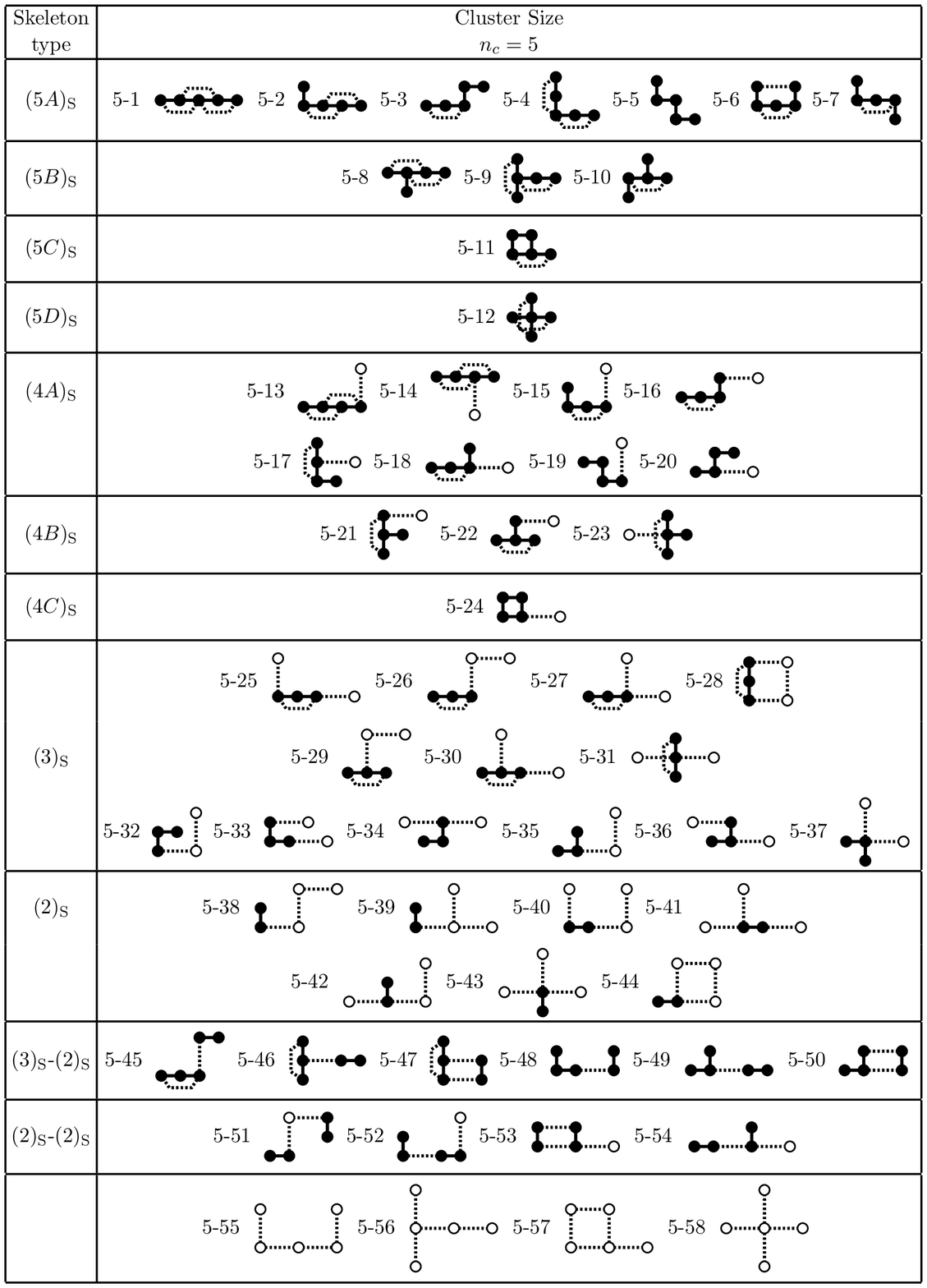}
\caption{\label{Fig4} The quintet types of the lopsided \J1-\J3 model.
The skeletons are emphasized. The skeleton type  is given in the first column.}
\end{figure*}

\subsection{\label{ss:IIIB}Skeletons}
The following definitions and properties for  the lopsided \J1-\J2 model also
apply
to the lopsided \J1-\J3 model. All that is necessary is to replace  ``\J2'' by
``\J3.''
\subsubsection{\label{sss:IIIB1}Definition of a skeleton}
Consider a cluster with at least one \J1 bond.
The skeleton of this  cluster consists of all the \J1 bonds in the cluster
together with
all the spins that are attached to these \J1 bonds.
The much weaker \J2 bonds, if any are present, are never included in a skeleton.
Some spins in the skeleton (called ``skeleton-spins'')  may be attached to \J2
bonds.
But even in such cases the \J2 bonds are not included in the skeleton.

The cluster types of the \J1-\J2 model were shown in Figs.~1 and 2 of the
preceding paper.\cite{\reftwo}
\Figures{1}{2} of the present paper are revised versions of the same figures.
The revisions emphasize the skeletons.
Skeleton-spins are represented by solid circles, in contrast to other spins
which are shown as
empty circles.
The \J1 bonds (always included in the skeleton) are represented by solid lines,
whereas the \J2 bonds (never  considered as parts of the skeleton) are shown as
dotted lines.

The cluster types of the \J1-\J3 model were shown earlier in Figs.~3 and 4 of
\Reftwo.
Revised versions of these figures, which emphasize the skeletons, are shown in
\Figs{3}{4} of the present paper.
The format is similar to  that in \Figs{1}{2}, except that dotted lines
represent \J3 bonds.

\subsubsection{\label{sss:IIIB2}Simple and compound skeletons}
Skeletons are either ``simple'' or ``compound.''
A skeleton is simple if and only if all the skeleton-spins are connected to each
other by a continuous path (or paths) of \J1 bonds. For a compound skeleton
there are
at least two skeleton-spins which are not connected by a continuous path of \J1
bonds.
Because these two spins are still in the same cluster, they must be connected by
at least
one continuous path of exchange bonds, but such a path must include at least one
\J2 bond.
Examples of compound skeletons are cluster type  4-12 in \Fig1,
and  cluster types  5-36 up to 5-41 in \Fig2.

A compound skeleton always consists of two or more ``fragments.''
All spins in the same fragment are connected by a continuous path (or paths)  of
\J1 bonds.
Any continuous exchange path between spins in different fragments  must include
at least one \J2 bond.

All cluster types in Figs.~\ref{Fig1}--\ref{Fig4}   contain no more than 5
spins.
Each fragment must contain at least two spins. Therefore, any compound skeleton
in these figures has two fragments, with  2 or 3 spins in each fragment.

\subsubsection{\label{sss:IIIB3}Skeleton types}
Consider simple skeletons first. By itself, any simple skeleton is identical to
some realization of one of the cluster types of the \J1 model.
This cluster type of  the \J1 model is chosen  as the type of the simple
skeleton.
The label for the skeleton type is also the same as the label  for the  cluster
type in  the  \J1  model (as defined in \Refone), except that it is surrounded by a parenthesis and is
followed
by a subscript ``$\Sk$.''
For example, a simple skeleton which is identical to a NN pair in the \J1 model
is a skeleton of type \skel{2}.
In the \J1 model the cluster type gives the complete set of \J1 bonds in the
cluster.
Similarly, the type of a simple-skeleton gives the complete set of \J1 bonds in
the simple skeleton.

Compound skeletons also are classified by types.
As a first step, each fragment is  assigned  a ``fragment  type.''
The procedure  is similar to the assignment of a type to  a simple skeleton.
The fragment type gives the complete set of \J1 bonds in the fragment.
For the compound skeleton as a whole, the  skeleton type is given by  the
sequence of the fragment types. The order in the sequence is arbitrary.
For example, the compound skeletons of quintet types 5-36, 5-37, and 5-38 in
\Fig2 are all composed of two fragments: a NN pair and a NN triplet.
These are cluster types $2$ and $3$ of  the  \J1 model (see Fig.~3 of ~\Refone).
The compound skeletons of the three mentioned cluster types in \Fig2 are
therefore  of type \skel{2}-\skel{3} or, alternatively, \skel{3}-\skel{2}.

Figures \ref{Fig1}--\ref{Fig4} are arranged so that cluster types with the same
skeleton type appear in the same row. The skeleton type is given in the left
column.
The cluster types in the bottom row of each of these figures have no \J1 bonds.
Therefore,  they have no skeletons.

\subsection{\label{ss:IIIC}Decorations}
\subsubsection{\label{sss:IIIC1} Decorations of mixed clusters}
Any mixed cluster has both a skeleton and a decoration.
The decoration  consists of all the exchange bonds that are not in the skeleton,
together with  all the spins that are not in the skeleton, if there are any such
spins.
For a mixed \J1-\J2 cluster type the decoration includes all the \J2 bonds, and
all the spins that are not attached to any \J1 bond.
Each  decoration-spin  must be attached to at least one \J2 bond;  otherwise it
would have been a single, and not a spin in a mixed \J1-\J2 cluster.
In Figs.~\ref{Fig1}--\ref{Fig4} the decoration spins are among the spins that
are represented  by empty circles.

Decorations of mixed \J1-\J2 cluster types may be divided into two classes:
\begin{enumerate}\listpar
\item  ``Spinless decorations'' (also called pure-bond decorations) contain no
spins.
They consist only of \J2 bonds. All the spins  of the  cluster are then in the
skeleton.
Therefore, all the \J2 bonds are between skeleton-spins.
Examples of spinless  decorations are those of cluster types
 3-2, 4-2, 4-3, 4-4, and 4-5 in \Fig1, and of  types  5-2 up to 5-11 in \Fig2.
\item   A ``spinned decoration'' (also called a spin-bond decoration) contains
at least one spin.
Any spin in such a decoration must be attached to at least one \J2 bond, but it
cannot be attached to a \J1 bond (otherwise, it would have been  a
skeleton-spin).
Examples of spinned decorations are those  in cluster types 3-3, and 4-6 up to
4-11 in \Fig1, and in cluster types from 5-12 up to 5-35 in \Fig2.
\end{enumerate}

Although all \J2 bonds of a mixed cluster are always parts of  the decoration,
there
is no requirement that all \J2 bonds of a mixed cluster must be  between
decoration-spins.
Some \J2 bonds can be between skeleton-spins, as in cluster types  4-7 and 4-8
(\Fig1).

Analogous definitions of spinless and spinned decorations apply to mixed
clusters of the lopsided \J1-\J3 model.
Their properties too are analogous to  those  of spinless and spinned
decorations of the lopsided \J1-\J2 model.

\subsubsection{\label{sss:IIIC2} Null decorations of  pure-\J1 clusters}
The concept of a ``decoration'' will also be used in connection with pure-\J1
clusters.
This usage may seem strange because all spins and all exchange bonds of a pure-
\J1
cluster are already included in its skeleton, i.e., the skeleton is the entire
cluster.
One may therefore reasonably say that a pure-\J1 cluster has no decoration.
The use of decorations in connection with pure-\J1 cluster types is just a
matter of convenience.

The ``existence'' of a decoration for a pure-\J1 cluster is merely a formality.
A ``null decoration'' is defined as a decoration that contains no spins and no
exchange bonds.
All  pure-\J1 clusters then have null decorations. Obviously, having a null
decoration is equivalent to having no decoration.

In summary,  there are three categories of decorations:
Spinless (pure-bond)) and spinned  (spin-bond)  decorations occur in mixed
clusters. Null   decorations occur in pure-\J1 clusters.

\subsection{\label{ss:IIID} Sizes of skeletons, decorations, and fragments}
The size of a skeleton is the number of spins in the skeleton, labeled as
$n_\Sk$.
The size $n_\D$ of a decoration is the number of spins in the decoration.
For null decorations and spinless  decorations, $n_\D{=}0$.
The size $n_c$ of any cluster type $c$  composed of a skeleton and a decoration
is
\begin{equation}
n_c = n_\Sk + n_\D.               \label{eq2}
\end{equation}
The size $n_\F$ of a fragment of a compound skeleton is the number of spins in
the fragment.
The size of the compound skeleton is the sum of $n_\F$ over all fragments in the
skeleton,
\begin{equation}
 n_\Sk =\sum n_\F.                \label{eq3}
\end{equation}

\section{\label{s:IV}MONO-SKELETON AND POLY-SKELETON FINE STRUCTURES IN THE
HIGH-FIELD PART OF THE SPECTRUM}
The concepts of skeletons and skeleton types bring insight into the physics of
the  FS in the high-field part of the spectrum.
The major cause of this FS is illustrated by the following example.

\subsection{\label{ss:IVA}Example of a mono-skeleton FS}
Several cluster types of  the \J1-\J2 model have a simple skeleton of type
\skel{2}.
The lower part of \Fig5 shows simulated spectra from all these cluster types.
The spectra are plotted against the primary reduced field $b_1$.
The parameters  for these simulations are: $\irJ2{=}0.028$, \Spin, $\inmax{=}5$,
and $T{=}0$.
For these parameters the high-field part of the spectrum starts at
$b_1{=}0.920$.
The lower part of \Fig5 shows that in this high-field part, spectral lines from
all the cluster types with a \skel{2} simple skeleton are very close to
$b_1{=}2, 4,\dots,10$.

The upper part of \Fig5 shows the spectra from all cluster types with a compound
skeleton  containing  a fragment of type \skel{2}.
All these cluster types  also  produce spectral lines very near to  $b_1{=}
2,4,\dots,10$.
Several of these cluster types (types  5-36, 5-37, and 5-38) also produce
additional lines near
$b_1{=} 7, 9,\dots,15$. These three cluster types have a compound skeleton of
type \skel{3}-\skel{2}.

In the (parent) \J1 model, the spectral lines of cluster type $2$ (NN pair)  are
exactly  at $b_1{=}2, 4,\dots,10$.
In the same model, cluster type $3$ (NN triplet) produces lines exactly at
$b_1{=}7, 9,\dots,15$.
The obvious interpretation of the high-field results in \Fig5 is, therefore,
that all cluster types of the \J1-\J2 model that contain a simple skeleton  or a
fragment  of
type \skel{2}  produce high-field lines at nearly the same magnetic fields as
those of lines produced by cluster type $2$ in the \J1 model.
The  additional  high-field lines in the upper part of \Fig5, from  cluster
types 5-36, 5-37, and 5-38,
are due to the \skel{3} fragment in the \skel{3}-\skel{2} compound skeleton.
The magnetic fields at these additional lines are nearly the same as those of
the
lines from the triplets (cluster type $3$) of the \J1 model.

The total MST spectrum is a statistically-weighted superposition of spectra from
all
cluster types  of the model.
There are $18$ cluster types  with a \skel{2} skeleton/fragment.
Together, they produce groups of very close lines near $b_1{=}2, 4,\dots,10$.
Line separations $\Delta b_2$ within each group are of order 1.
The  very close lines  in each  group   may be viewed as a FS splitting of a
single line from   NN pairs (cluster type $2$) in the \J1 model.
This  FS,  will be called the \skel{2} ``mono-skeleton''  FS.
The dark lines in \Fig5 represent the spectral  lines of   this mono-skeleton
FS.

\begin{figure}\includegraphics[scale=1]{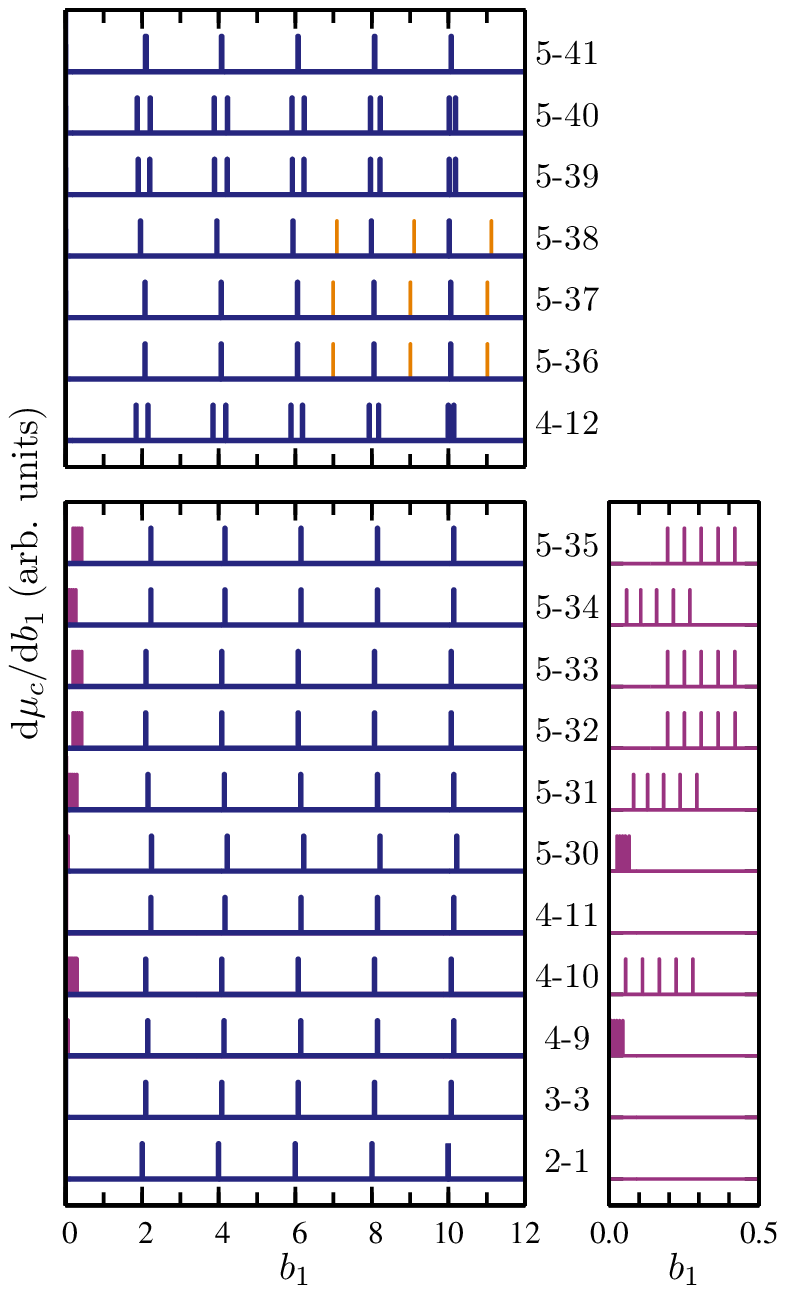}
\caption{\label{Fig5}
The spectra of all cluster types that contribute to the \skel{2} mono-skeleton
FS in the \J1-\J2 model.
Lines in the  \skel{2} mono-skeleton FS are darker (dark blue).
The lower part of the figure shows the spectra from all cluster types  with a
simple skeleton of type \skel{2}.
The panel to the right of the  lower part gives an expanded view of the low-
field spectra.
The upper part of the figure shows the spectra  from all cluster types with a
compound skeleton containing a fragment of type \skel{2}.
The labels for the cluster types follow \Figs{1}{2}.
The abscissa  is the primary reduced magnetic field $b_1$.
All line intensities are chosen to be equal.}
\end{figure}

\begin{figure}\includegraphics[scale=1]{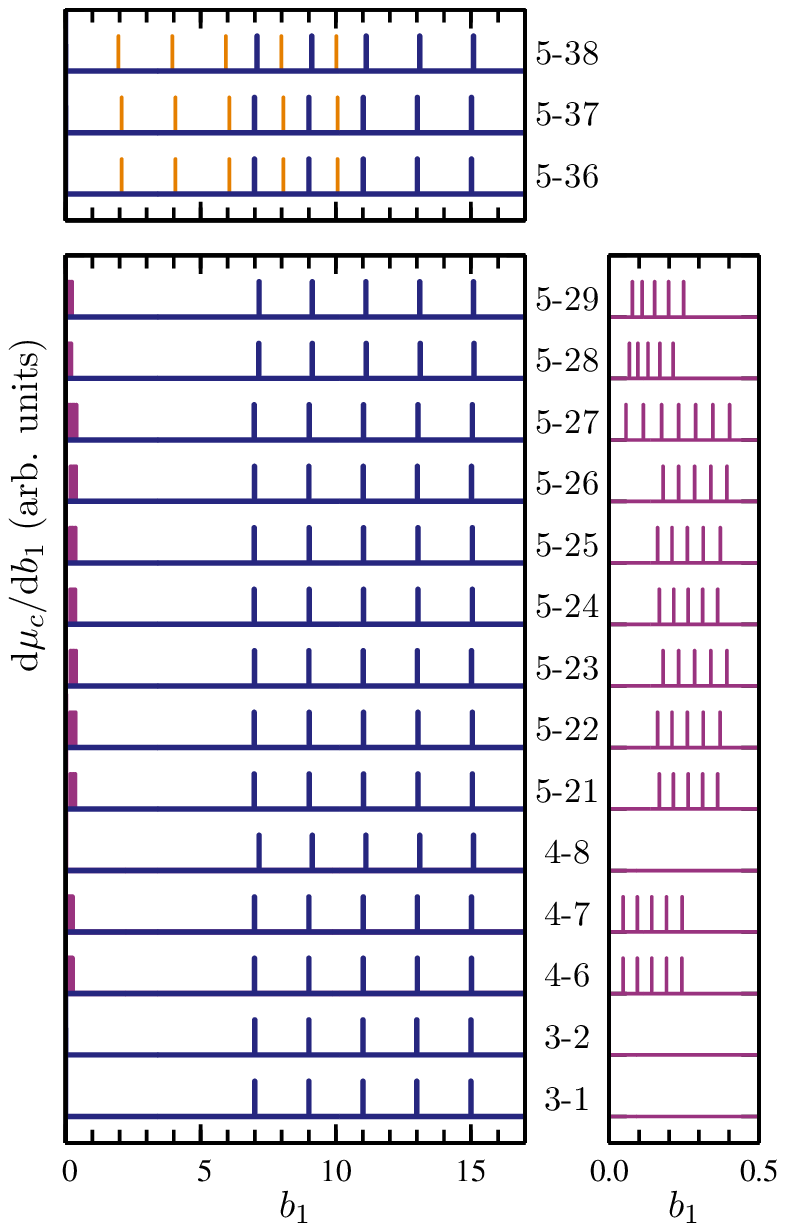}
\caption{\label{Fig6}
The spectra of all cluster types that contribute to the \skel3 mono-skeleton FS
in the \J1-\J2 model.
The format is similar to that in \Fig5, except that the darker (dark blue) lines
are now the lines in the \skel{3} mono-skeleton FS.
The labels for the cluster types follow \Figs{1}{2}.}
\end{figure}

\begin{figure}\includegraphics[scale=1]{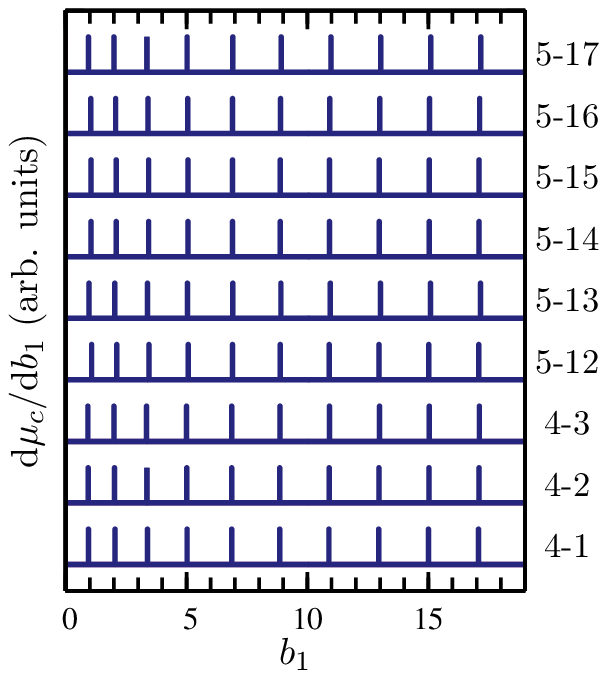}
\caption{\label{Fig7}
The \skel{4A} mono-skeleton spectra of the \J1-\J2 model.
The labels for the cluster types follow \Figs{1}{2}.}
\end{figure}

\begin{figure}\includegraphics[scale=1]{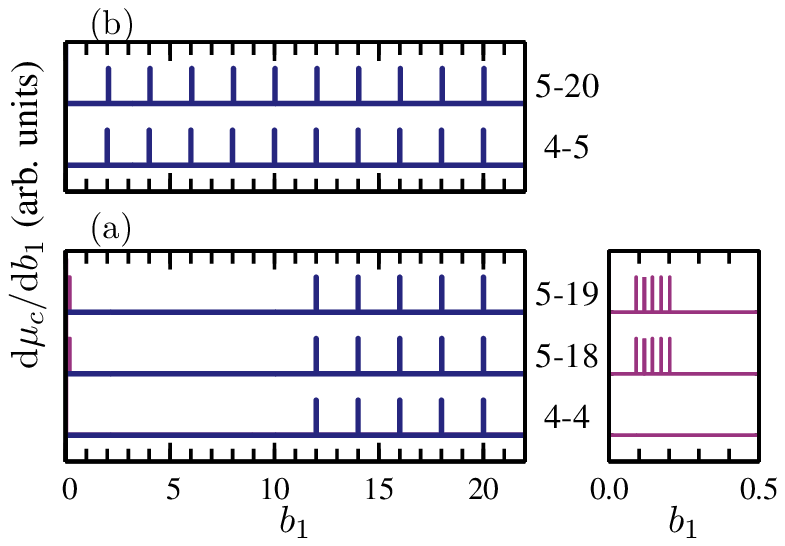}
\caption{\label{Fig8}
(a) The spectra of all cluster types that contribute to the \skel{4B}
mono-skeleton FS in the \J1-\J2 model.
Only the darker (dark blue) lines are in the \skel{4B} mono-skeleton FS.
The panel on the right gives an expanded view of the low-field spectra.
(b) The \skel{4C} mono-skeleton spectra of the \J1-\J2 model.}
\end{figure}

\begin{figure}\includegraphics[scale=1]{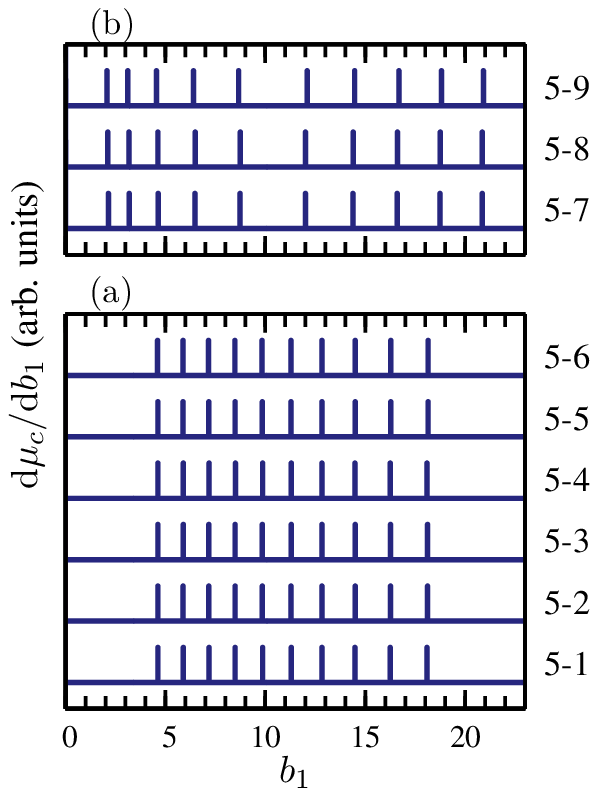}
\caption{\label{Fig9}
(a) The \skel{5A} mono-skeleton spectra of the \J1-\J2 model.
(b) The \skel{5B} mono-skeleton spectra of the \J1-\J2 model.}
\end{figure}

\subsection{\label{ss:IVB}Other mono-skeleton FS's in the \J1-\J2 model}
The results below are based on simulations of spectra from various cluster types
of the \J1-\J2 model. Once again the parameters are: $J_2{/}J_1{=}0.028$, \Spin,
$n_\mathrm{max}{=}5$, and $T{=}0$. The discussion focuses on mono-skeleton FS's.
All such FS's are in the high-field part of the spectrum.

The bottom part of \Fig6 shows the spectra from all cluster types  with  a
simple skeleton of type \skel{3}.
All  the high-field lines are very near $b_1{=}7, 9,\dots,15$.
The upper part of \Fig6 shows the spectra from the three cluster types that
contain a \skel{3} fragment.
The same three cluster types already appeared in \Fig5.
They contain both a \skel{3} fragment and a \skel{2} fragment.
The \skel{3} fragment produces lines very near $b_1{=} 7, 9,\dots,15$, and
the \skel{2} fragment produces lines very close to close to
$b_1{=}2,4,\dots,10$.

The  \skel{3} mono-skeleton FS is  produced by the $17$ cluster types with
\skel{3} skeletons/fragments.
This FS consists of groups of very close lines near $b_1{=} 7, 9,\dots 15$.
Line separations, $\Delta b_2$, in each group are of order 1.
Each group  of very close lines may be viewed as a  FS splitting of one of  the
triplet  lines (cluster type $3$) in  the \J1 model.
In \Fig6, the  \skel{3} mono-skeleton FS is represented  by dark lines.

All cluster types in \Figs{1}{2} have sizes $n_c{\leq}5$.
The only possible fragment types are \skel{2} and \skel{3}.
The spectra from all cluster types containing such fragments  have already been
included in \Figs{5}{6}.
The  only high-field   lines that are not included in \Figs{5}{6}  are from
cluster types with simple skeletons of sizes $ n_\Sk{=}4$ or $5$.

\Figure7 shows the spectra from all cluster types  with a \skel{4A} skeleton.
In the parent \J1 model the $4A$ quartets produce lines at $b_1{=}0.950,
2.041,3.389,\dots$
(see Table II of \Refone). The combined spectrum from all nine cluster types in
\Fig7 consists of
groups of several very close lines near each of these values of $b_1$.
This is the \skel{4A} mono-skeleton FS of the lopsided \J1-\J2  model.

\Figure8(a) shows the spectra from the three cluster types with a \skel{4B}
skeleton.
The high-field lines from these cluster types are all very close to each other.
The \skel{4B} mono-skeleton FS
consists only of these high-field lines (shown darker).
The two cluster types in \Fig8(b) are responsible for the \skel{4C}
mono-skeleton FS.

\Figure9(a) shows the spectral lines from the six cluster types in \Fig2 that
have a \skel{5A} skeleton. The \skel{5A} mono-skeleton FS includes all these
lines.
The \skel{5B} mono-skeleton FS is produced by the three cluster types in
\Fig9(b).

There is only one cluster type with a \skel{5C} skeleton.
Its spectrum (not shown) is very similar to that of cluster type $5C$ in the
(parent)   \J1 model.
All the lines are in the high-field part of the spectrum.
Because there is only one cluster  type, a mono-skeleton FS does not develop.
Similar remarks apply to the one cluster type with a \skel{5D} skeleton.

\subsection{\label{ss:IVC}Poly-skeleton FS in the high-field part of the
spectrum}
A poly-skeleton FS occurs whenever different mono-skeleton FS's, associated with
different skeleton/fragment types, overlap in the same region of the magnetic
field.
For example, Table~I of \Refone\ shows that in the parent \J1 model all five
lines from cluster type $2$ (at $b_1{=} 2, 4,\dots,10$) coincide with the first
five
lines from cluster type  $4C$.
In the lopsided \J1-\J2 model, each  group  of very close lines  in the \skel2
mono-skeleton FS will overlap a group of very close lines  in the  \skel{4C}
mono-skeleton FS.

In the example just given, spectral lines in the \J1 model from  two different
cluster types of that model were at exactly the same fields.
Exact coincidence of lines from different cluster types of the \J1 model,
however, is not the only
possible origin of a poly-skeleton FS.
A close proximity of lines from different cluster types of the \J1 model is
another possibility.
For example, in the \J1 model  cluster type $4A$ has a line at $b_1{=}2.041$.
This line is quite close to the line at $b_1{=}2$ from cluster type $2$.
Unless \rJ2 is extremely small, the \skel2 and \skel{4A} mono-skeleton FS's, by
themselves,
will generate a poly-skeleton FS near $b_1{=}2$.
However, as already noted, cluster type $4C$ of the \J1 model also has a line at
$b_1{=}2$,
so that the \skel{4C} mono-skeleton FS will also contribute to this
poly-skeleton FS.
Finally, cluster type $5B$ of the \J1 model has a line at $b_1{=}2.166$, which
is not too far from $2$.
Depending  on the ratio \rJ2, the \skel{5B} mono-skeleton FS may also contribute
to the
poly-skeleton FS near $b_1{=}2$.

\begin{figure}\includegraphics[scale=1]{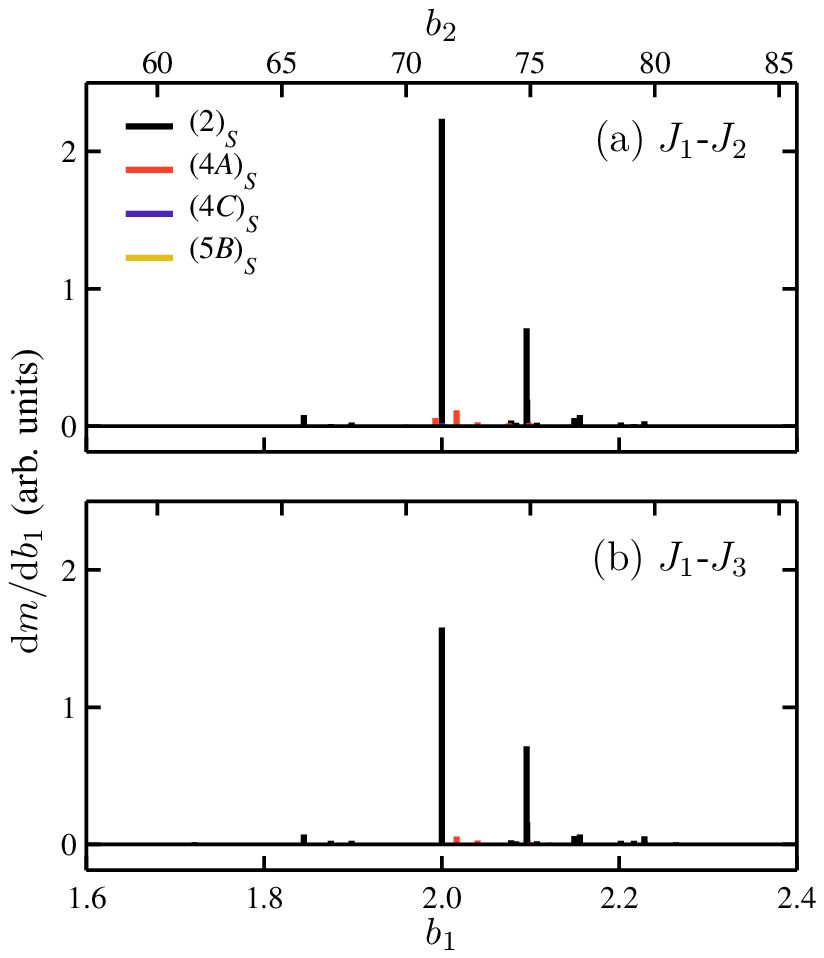}
\caption{\label{Fig10}
(a) The full spectrum of the \J1-\J2 model near $b_1{=}2$, calculated for
$x{=}0.16$, $\irJ2{=}0.028$, \Spin,  and $T{=}0$.
Only lines belonging to the \skel{2}, \skel{4A}, \skel{4C} and \skel{5B} mono-
skeleton
spectra are in this field range.   The \skel{4C} and \skel{5B} lines are
invisible in
this figure because their intensities are very low.
(b) The analogous results for the \J1-\J3  model.
The lower and upper abscissa scales are for the primary and secondary reduced
field, $b_1$ and $b_2$, respectively.}
\end{figure}

\begin{figure}\includegraphics[scale=1]{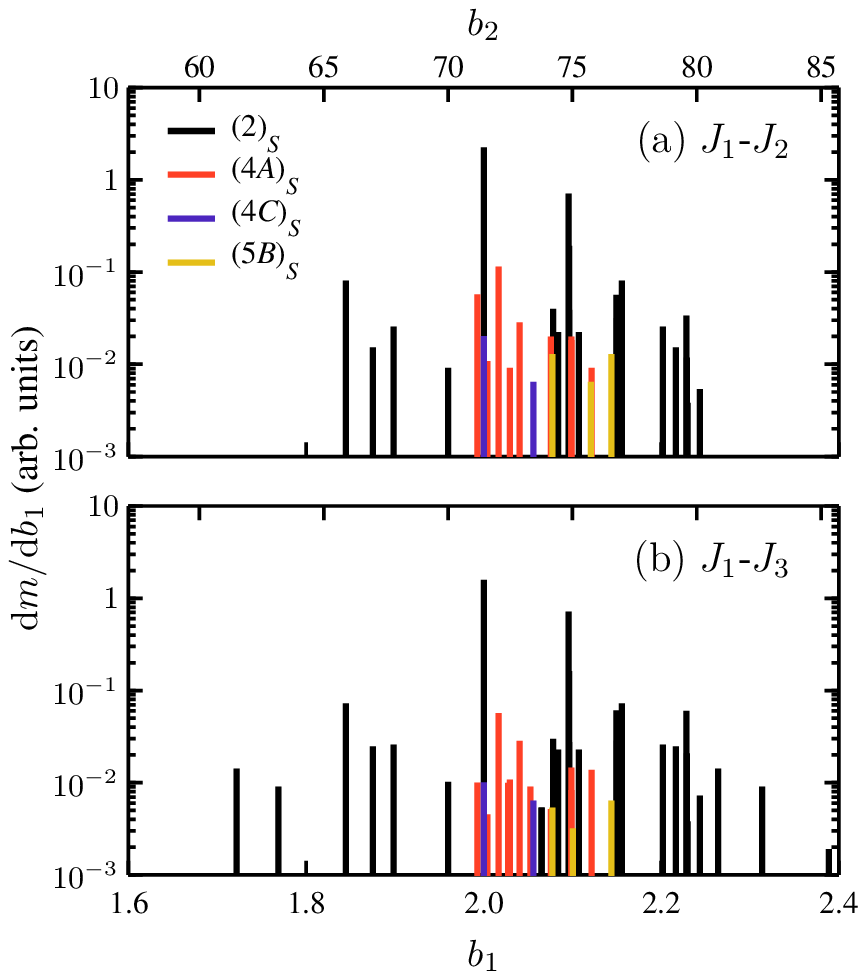}
\caption{\label{Fig11}
The spectra in  Figs. \ref{Fig10}(a) and \ref{Fig10}(b), replotted using a
log scale for the line intensities.}
\end{figure}

\Figure{10}(a) shows the full spectrum of the \J1-\J2 model near $b_1{=}2$,
calculated for $x{=}0.16$ with    $\irJ2{=}0.028$.
Only  strong lines are visible in this figure. The weak lines are seen clearly
in
\Fig{11}(a)  which uses a log scale for the line intensities.
As expected, the poly-skeleton spectrum in this field range is composed of four
mono-skeleton FS's: \skel{2}, \skel{4A}, \skel{4C}, and \skel{5B}.
The relative intensities of spectral lines are always governed by cluster
statistics.
In Figs.~\ref{Fig10}(a) and \ref{Fig11}(a) the strongest lines are from the
\skel{2} mono-skeleton FS.
As discussed later, the two most intense lines in this portion of the spectrum
are very useful for an experimental determination of \JJ2.

\subsection{\label{ss:IVD}Physical interpretation}
\subsubsection{\label{ss:IVD1}Changes of exchange and Zeeman energies at MST's}
Consider any cluster type other than a single. A MST, and the
associated
spectral line, occur at a magnetic fields $B$ where a crossing of two energy
levels changes the ground state.
The energy $E$ is the sum of  the exchange energy $E_\mathrm{ex}$ and Zeeman
energy $E_\mathrm{Z}$.
At any MST  the difference $\Delta E_\mathrm{ex}$ between the exchange energies
of the two levels
that cross is equal and opposite to the difference $\Delta E_\mathrm{Z}$ between
their Zeeman energies.

At any MST, the  $z$ component (along $\mathbf{B}$) of the ground-state total
spin changes by one.
For the  energy levels that cross, $\Delta E_\mathrm{Z}$ is therefore
proportional to $B$, and so is $\Delta E_\mathrm{ex}$.
Thus, at any high-field spectral line (above the gap) the changes $\Delta
E_\mathrm{z}$ and
$\Delta E_\mathrm{ex}$  are larger than at any low-field line.
The reduced fields $b_1$ at the high-field lines are of order $1$ or larger.
The change $|\Delta E_\mathrm{Z}|{=}|\Delta E_\mathrm{ex}|$ at any  high-field
line is therefore of order $|J_1|$ or larger.

\subsubsection{\label{ss:IVD2}Physical  interpretation of  a  mono-skeleton fine
structure}
The exchange energy of a cluster is due to all the exchange bonds.
Each bond  is between  two spins.
The abrupt change $\Delta E_\mathrm{ex}$ at any MST is caused by an abrupt
change in the
relative orientations of spins that are on opposite sides of one, or more,
exchange bonds.
To produce a change $\Delta E_\mathrm{ex}$ of order $|J_1|$, or larger, at any
of the high-field
lines, one or more of these bonds must be a \J1 bond.
The small number of \J2 bonds in any of the clusters in \Figs{1}{2} cannot
produce a sufficiently
large $\Delta E_\mathrm{ex}$ when the ratio \rJ2 is so small that there is a gap
in the spectrum.
All the \J1 bonds are in the skeleton. Therefore, at any high-field line there
is an abrupt change
in the relative orientations of skeleton-spins.

All skeleton-spins  are attached to \J1 bonds. However, some skeleton-spins may
also be attached to \J2 bonds. At some high-field lines,  simultaneous abrupt
changes in the energies of both \J1 bonds and \J2 bonds may occur.
However, the major contribution to $\Delta E_\mathrm{ex}$ is from one or more of
the strong \J1 bonds.

The arrangement of the strong  \J1 bonds in a simple skeleton is fully specified
by the skeleton type.
The arrangement of the \J1 bonds in a fragment of the same type is identical.
Therefore one expects that spectral lines produced by all simple skeletons and
fragments of one type will
involve nearly equal changes in the energy  of the \J1 bonds.
Because the energy change from  the \J1 bonds is the major contribution to
$\Delta E_\mathrm{ex}$,
and because the magnetic field $B$ at  the MST is proportional to $\Delta
E_\mathrm{ex}$, all
lines from skeletons and fragments of the same type should occur at
approximately the same field $B$.
The small differences, $\Delta B$, are due to differences in the minor
contribution  to
$\Delta E_\mathrm{ex}$ from the \J2 bonds.
Different cluster types with the same skeleton/fragment type have different
decorations,
with different arrangements of \J2 bonds.
It is these different  decorations that are responsible for the small line
separations,
$\Delta b_2{\sim}1$,  in the mono-skeleton FS.

\section{\label{s:V}LOW-FIELD SPECTRA}
\subsection{\label{ss:VA}General conclusions from numerical simulations}
MST's in the low-field part of the spectrum involve only the weak exchange
bonds, i.e.,
\J2 or \J3 bonds. The strong \J1 bonds are not involved.
Spins coupled by weak bonds occur in pure-\J2 or pure-\J3  cluster types, and in
mixed cluster types.
The low-field lines from pure-\J2  and pure-\J3 cluster types have already been
discussed in Sec.~\ref{ss:IIB}.
The remaining discussion therefore focuses on the mixed cluster types.
A  mixed cluster always has both a skeleton and a decoration. The decoration is
either spinned or spinless.

Numerically-simulated  spectra from many cluster types of the lopsided \J1-\J2
model have already been presented in Figs.~\ref{Fig5}--\ref{Fig9}. The panels on
the
right of the lower parts of Figs.~\ref{Fig5}, \ref{Fig6}, and \ref{Fig8}(a)
show the low-field spectra from all mixed cluster types of this model that have
sizes $n_c{\leq}5$.
These low-field spectra lead to the following general conclusions:
\begin{enumerate}\listpar
\item  Some cluster types with spinned decorations produce low-field lines, but
others  do not.
The cluster types that do, and those that do not, will be listed in
Sec.~\ref{ss:VB}.
The physics behind these lists will be discussed in the same section.
\item Cluster types with spinless decorations do not produce low-field lines if
the cluster
size is $n_c{\leq}5$. Therefore, none of the spinless decorations in \Figs{1}{2}
produce low-field lines.
However, it will be shown in Sec.~\ref{ss:VC} that spinless decorations in some
cluster types of sizes $n_c{>}5$ can produce low-field lines.
\end{enumerate}
The following discussion is for the lopsided \J1-\J2 model. Analogous results
apply to the lopsided \J1-\J3 model.

\subsection{\label{ss:VB}Low-field MST's from spinned decorations of the
lopsided \J1-\J2 model}
Any decoration-spin is  attached to one or more \J2 bonds, but not to any \J1
bond.
A \J2 bond which is attached to a  decoration-spin must also be attached either
to
another decoration-spin or to a skeleton-spin. The two cases are considered
separately.

\subsubsection{\label{sss:VB1}\J2 bond between two decoration-spins}
Any cluster type with  a \J2 bond between two decoration-spins always
produces one or more low-field spectral lines.
Examples  are the low-field lines of  cluster types 4-10, 5-23, 5-26,
5-27, 5-29, and 5-31 up to 5-35.
With the exception of cluster type 5-27, which will be discussed separately,
the relevant \J2 bond is not frustrated.
That is, the two decoration-spins attached to the \J2 bond are antiparallel at
$B{=}0$.
Low-field MST's, in reduced fields $b_2$ of order  $1$,
correspond  to a stepwise alignment of these decoration-spins.
At the completion of  these low-field MST's,  the decoration-spins attached to
the \J2 bond are parallel.

The preceding argument does not apply to cluster type 5-27.
The three \J2 bonds that are connected to the two decoration-spins cannot be all
satisfied at $B{=}0$.
The strong \J1 bonds lead to a parallel alignment of the two
end-spins in  the skeleton.
To satisfy the \J2 bonds between these two end-spins and the two
decoration-spins, the two
decoration-spins must be parallel to each other. But then the \J2 bond between
the two
decoration-spins will not be satisfied.
Thus, frustration of at least one of these three \J2 bonds is inevitable.
Because of this frustration, the spin configuration in the zero-field
ground-state of cluster type 5-27 is not immediately obvious.

The following physical picture for cluster type 5-27  is based, in part, on the
results in \Fig6.
In the zero-field ground state  the net spin of the skeleton alone is
$S_\Sk(0){=}5/2$;
the two parallel end-spins of the skeleton are  antiparallel to the
middle spin.
The two decoration-spins, in this  zero-field ground state, have a net spin
$S_\D(0){=}2$.
The two \J2 bonds between the skeleton's end-spins and the decoration-spins
cause  the net  spin of the decoration to be antiparallel to net spin of the
skeleton.
The zero-field ground state of the entire cluster therefore has a total spin
$S_c(0){=}5{/}2{-}2{=}1{/}2$.

Cluster type 5-27 has seven MST's in low fields.
At each of these MST's the total spin $S_c$ of the cluster's ground state
increases by  one unit.
At the completion of these seven MST's, the decoration-spins and the end-spins
of  the skeleton are all parallel to each other.
The decoration spin then has a magnitude   $S_\D{=}5$, and
it is parallel to the skeleton's net spin whose magnitude is still
$S_\Sk{=}5{/}2$.
Thus, at the completion of the seven low-field MST's, $S_c{=}15{/}2$ for the
entire cluster. Five
 MST's above the gap in the spectrum, increase
$S_\Sk$  from $5{/}2$ to $15{/}2$, with no change in $S_\D$
which  remains saturated at $5$. All spins of the cluster are then
parallel, so that
$S_c{=}25{/}2$.

\subsubsection{\label{sss:VB2}  \J2 bonds  between decoration-spins and
skeleton-spins}
The only spinned decorations that are not covered by the preceding discussion
are those in which all \J2 bonds  that are attached to decoration-spins  are
also
attached to skeleton-spins.
Among the many cluster types with such spinned decorations, some produce
low-field lines but others do not.
The  number of low-field lines (zero if there are no low-field lines) can be
obtained as follows.

The total spin $S_c$ of the cluster's ground state changes by unity at each MST.
Therefore, the total number of MST's, $\NTOT$, for a cluster type $c$ with size
$n_c$  is given by
\begin{equation}
                \NTOT=n_cS - S_c(0),            \label{eq4}
\end{equation}
where $n_cS$  and $S_c(0)$ are the saturation and zero-field values of $S_c$,
respectively.
Similarly, the number of MST's in the  high-field region, $\NHF$, is
\begin{equation}
               \NHF  = n_\Sk S - S_\Sk(0),            \label{eq5}
\end{equation}
where $n_\Sk S$  and $S_\Sk(0)$ are the saturation value and the zero-field
value of the skeleton's ground-state spin, respectively.
The number $\NLF$ of low-field MST's is
\begin{equation}
               \NLF=\NTOT  - \NHF.            \label{eq6}
\end{equation}

For the cluster types under consideration, $S_\Sk(0)$ and $S_c(0)$, can be
obtained by inspection because in the zero-field ground state none of the
\J1 bonds (all in the skeleton) are frustrated, and none of those \J2 bonds
that are between skeleton-spins and decoration-spins are frustrated.
Using $S_\Sk(0)$ and $S_c(0)$, Eqs.~(\ref{eq4})--(\ref{eq6}) then give $\NTOT$,
$\NHF$, and $\NLF$.
Table~\ref{t:I} gives the results for all the relevant cluster types, assuming
that \Spin.
All the values of $\NLF$ in this table agree with the simulations
[see the low-field results in Figs.~\ref{Fig5}, \ref{Fig6}, and \ref{Fig8}(a)].

\begin{table}
\caption{\label{t:I}
Properties of  those cluster types of the lopsided \J1-\J2 model for which all
\J2 bonds that are attached to decoration-spins are also attached to
skeleton-spins.
It is assumed that \Spin. $\NTOT$  is the total number of MST's.
$\NHF$ is the number of high-field MST's (above the gap), and $\NLF$  is the
number of low-field MST's.
The saturation value  of the cluster's spin is $n_cS$.
For the skeleton alone, the saturation value is  $n_\Sk S$.
In the zero-field ground state, the cluster's spin is $S_c(0)$, and the
skeleton's spin is $S_\Sk(0)$.}
\begin{ruledtabular}
\begin{tabular}{|c|c|c|c|c|c|c|c|}
Cluster&&&&&&&\\
Type&$n_cS$&$S_c(0)$&$\NTOT$&$n_\Sk S$&$S_\Sk(0)$&$\NHF$&$\NLF$\\\hline
3-3&15/2&5/2&5&5&0&5&0\\
4-6&10&0&10&15/2&5/2&5&5\\
4-7&10&0&10&15/2&5/2&5&5\\
4-8&10&5&5&15/2&5/2&5&0\\
4-9&10&0&10&5&0&5&5\\
4-11&10&5&5&5&0&5&0\\
5-12&25/2&5/2&10&10&0&10&0\\
5-13&25/2&5/2&10&10&0&10&0\\
5-14&25/2&5/2&10&10&0&10&0\\
5-15&25/2&5/2&10&10&0&10&0\\
5-16&25/2&5/2&10&10&0&10&0\\
5-17&25/2&5/2&10&10&0&10&0\\
5-18&25/2&5/2&10&10&5&5&5\\
5-19&25/2&5/2&10&10&5&5&5\\
5-20&25/2&5/2&10&10&0&10&0\\
5-21&25/2&5/2&10&15/2&5/2&5&5\\
5-22&25/2&5/2&10&15/2&5/2&5&5\\
5-24&25/2&5/2&10&15/2&5/2&5&5\\
5-25&25/2&5/2&10&15/2&5/2&5&5\\
5-28&25/2&5/2&10&15/2&5/2&5&5\\
5-30&25/2&5/2&10&10&0&5&5\\
5-39&25/2&5/2&10&10&0&10&0\\
5-40&25/2&5/2&10&10&0&10&0\\
5-41&25/2&5/2&10&10&0&10&0
\end{tabular}
\end{ruledtabular}
\end{table}

\subsection{\label{ss:VC}Physical picture for the  absence  or presence of low-field
MST's from spinless decorations}
When the decoration is spinless, all spins of  the cluster are skeleton-spins.
The skeleton is either simple or compound. The two cases are considered
separately.
\subsubsection{\label{sss:VC1}Simple skeletons}
A cluster with a simple skeleton and a spinless decoration does not produce
low-field lines.
The reasons are as follows.
Any two spins in the cluster are connected by a continuous path (or paths) of
\J1 bonds.
For the cluster sizes considered here, any such a ``\J1 bridging path''  is
relatively short, i.e., the number of \J1 bonds in  the path is of order $1$.
An energy of order $|J_1|$ is required to produce an abrupt change of the
relative
orientations of spins in such a bridging path.

Because the decoration is spinless, any weak bond (\J2 bond) is between two
skeleton-spins.
These two skeleton-spins are also connected by  one or more \J1 bridging paths.
(For examples see cluster types 4-2 up to 4-5, and 5-2 up to 5-11, in
\Figs{1}{2}).
If a low-field MST occurs, it must arise from abrupt changes in the relative
orientations of skeleton-spins that are coupled by  \J2 bonds.
But such abrupt changes  will always be accompanied by an abrupt energy change
of
order \J1 from at least one \J1 bridging path. An energy change of order \J1 is
incompatible with the assumption that the MST is in the low-field part of the
spectrum.

\subsubsection{\label{sss:VC2}Compound skeletons}
The preceding arguments for a spinless decoration may fail  if the skeleton is
compound rather than simple.
The reason is that at least one \J2 bond is between spins in different
fragments. Such spins are not connected by a \J1 bridging path.

In the zero-field ground state of the cluster, each fragment has a net spin,
say,  $\mathbf{S}_{\F1}$ for one fragment and $\mathbf{S}_{\F2}$ for the other.
If neither  of these fragment-spins is zero at $B{=}0$ then the \J2 bond(s)
between the fragments  will cause $\mathbf{S}_{\F1}$ and $\mathbf{S}_{\F2}$ to
be
antiparallel in zero field.
In the low-field region ($b_2$  of order $1$), a series of MST's will then take
place.
This low-field series will  end when $\mathbf{S}_{\F1}$ and $\mathbf{S}_{\F2}$
become parallel.

As an example, consider a cluster type of size $n_c{=} 6$ with a
\skel{3}-\skel{3} compound skeleton. The two \skel{3} fragments are coupled
by only one \J2 bond.
The spinless decoration consists of  the lone \J2 bond linking the fragments.
The spin arrangement in the zero-field ground state is represented by
$({+}{-}{+})\cdots({-}{+}{-})$, where the dotted line is the \J2 bond between
the fragments.
Each fragment then has a spin $S_\F{=}5{/}2$.
The antiferromagnetic \J2 bond leads to a zero net spin for the entire cluster.
Five MST's will occur in low-fields, ending with  the spin arrangement
 $({+}{-}{+})\cdots({+}{-}{+})$. The cluster's net spin is then $5$.
Additional  MST's that involve the \J1 bonds within the fragments will occur in
the high-field region.

The preceding example involved a cluster type of size $n_c{>}5$.
When $n_c{\le}5$, at least one fragment of the compound skeleton  is a \skel2
fragment.
In the zero-field ground state,  this fragment has  net spin  $S_{\F1}{=0}$.
The \skel2 fragment is therefore not responsive to a magnetic field in low
fields.
The fragment becomes responsive  only when $S_{\F1}$ can change from zero.
The first such possibility is when $b_1$  is near $2$,    where  $S_{\F1}$
can change from zero to $1$.
Regardless of whether the other fragment is of type \skel2 or of type \skel3,
there are no MST's in the low-field part of the spectrum, which  is well below
$b_1{=}2$.

In summary, in order for a spinless decoration to produce MST's in the  low-
field part of the spectrum,
it is necessary that: 1) the skeleton is compound, and 2) the cluster size is
$n_c{>}5$.
These necessary conditions are not always sufficient.

\section{\label{s:VI} MEASUREMENT OF THE SECOND-LARGEST EXCHANGE CONSTANT}
It is assumed that the second-largest exchange constant \JJ2 is either \J2 or
\J3.
Two issues of interest to experimentalists are:
1) What features of a measured spectrum are most useful for determining the
magnitude of \JJ2?
2) Is there a practical way of deciding whether \JJ2 is \J2 or \J3?
\subsection{\label{ss:VIA}Determination of \JJ2 from the low-field spectrum}
For both the \J1-\J2 and \J1-\J3  models, cluster type 2-2 corresponds to a pair
in
which the two spins are coupled by \JJ2.
When the models are lopsided, these 2-2 pairs produce the strongest lines in the
low-field
part of the spectrum, at least up to $x{=}0.20$
(see Fig.~13 of~II).
The magnetic fields at these strongest low-field lines yield \JJ2 via Eqs.~(3b)
and (6b) of~II.

To resolve the 2-2 lines, the line width $\delta B$  must be smaller than the
field separation
between adjacent lines, $\Delta B {=} 2|J^{(2)}|/g\muB$,  or $\Delta b_2{=}2$.
Thermal broadening, proportional to $T$, places a lower limit on the line width
$\delta B$
(see, e.g., \Review).
For a lopsided model,  \JJ2 can be so small that very low temperatures are
required to resolve the 2-2 lines.

\subsection{\label{ss:VIB}Determination of \JJ2 from high-field FS}
The high-field FS near $b_1{=}2, 4,\dots,10$ provides an alternative route for
determining \JJ2.
Figures~\ref{Fig10}(a) and \ref{Fig10}(b) illustrate the FS near $b_1{=}2$ in
the \J1-\J2 and
\J1-\J3 models, respectively.
In both models the 2-1 line, from \J1 pairs, is the strongest.
The second-strongest line is from 3-3 triplets, each of which consists of a  \J1
pair attached to
a third spin by a \JJ2 bond.
Both the 2-1 and the 3-3 lines belong to the \skel{2}  mono-skeleton FS.

\begin{figure}\includegraphics[scale=1]{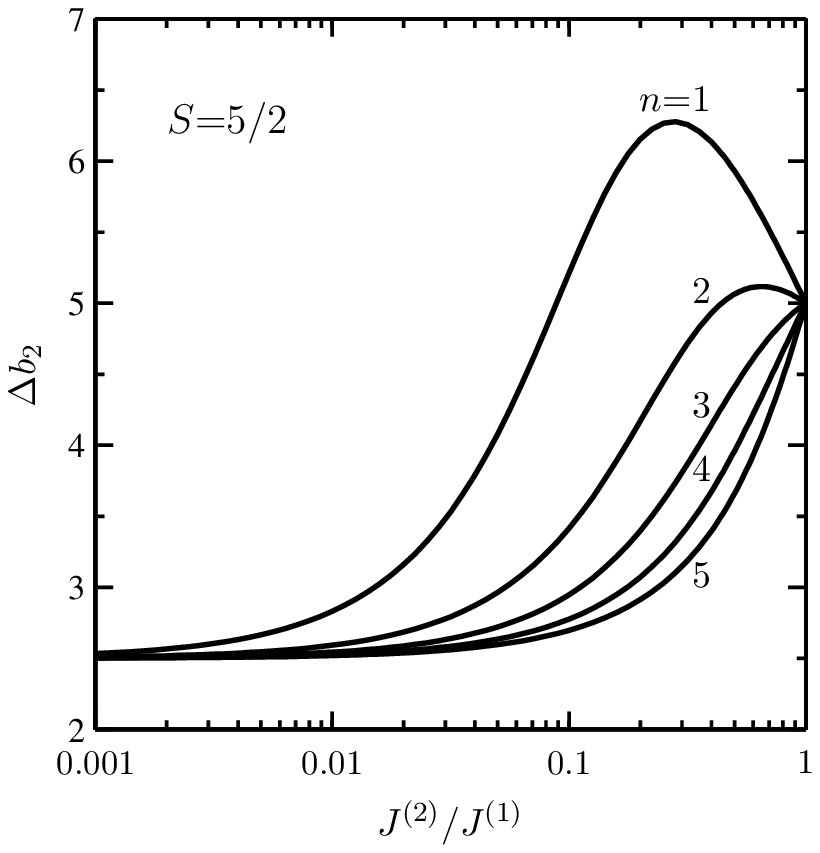}
\caption{\label{Fig12}
The separation $\Delta b_2$ between the secondary reduced fields $b_2$
at the 3-3 and 2-1 lines that are near $b_1{=}2n$.
These results, for $n{=}1,2,\dots5$, are plotted as a function of \rJJ.
The curves for $\irJJ{\le}1$ apply to both the lopsided \J1-\J2 and the lopsided
\J1-\J3 models.
(Adapted from Refs. \onlinecite{\review} and \onlinecite{Vu92thesis}).}
\end{figure}

The 2-1 and 3-3 lines are the strongest two lines not only in the FS near
$b_1{=}2$, but also
in the FS near each of the values $b_1{=}2n$, where $n{=}1, 2,\dots,5$.
(The groups of very close lines near $b_1{=} 2, 4$, and $6$ are shown in Fig.~12
of~\Reftwo.)
The separation $\Delta B$ between the 2-1 and 3-3  lines was calculated earlier
for all values
of $n$.\cite{\review,Vu92thesis}
The results, as a function of \rJJ,  are reproduced in \Fig{12}.
The ordinate is the field separation in reduced units, $\Delta b_2 = g\muB
\Delta B/|J^{(2)}|$.
\Figure{12} indicates that $\Delta b_2$ is largest for $n{=}1$, i.e., near the
first MST from pure \J1 pairs.
The measured separation between the 2-1 and 3-3  lines close to $b_1{=}2$,
together with the results in \Fig{12}, can be use to determine \JJ2 .

\subsection{\label{ss:VIC}Difficulty of identifying \JJ2 as \J2 or \J3}
The spectra and magnetization curves obtained from the \J1-\J2 and \J1-\J3
models are not identical.
In principle, it should be possible to distinguish between \J2 or \J3 by comparing
the
measured spectrum and/or magnetization curve with simulations based on
the two models.\cite{\review,\refthree,Bindi98prl}
Because the simulations assume a random distribution of the  magnetic ions
over the cation sites, conclusions concerning the identity of \JJ2 that are
based on such comparisons hinge on this assumption.
A random distribution has been found in many materials, but deviations
(sometimes large) have also been observed (see \Review).

In practice, even if the distribution is random it may be difficult to
distinguish between \J2 and \J3.
The basic reason is that the number of 2nd and 3rd neighbors is the same,
namely, $4$.
Although spectra calculated from the \J1-\J2 and \J1-\J3 models are not
identical,
the differences are often quite subtle.
Many of these differences are in the low-intensity lines (compare, for example,
the two parts of \Fig{11}).
Obviously, these are just the lines that are the most difficult to observe and
measure accurately.

The comparison of the magnetization curve, rather than the spectrum, with
simulations may be more promising.
The change of $M$ in a field interval that includes many spectral lines is the
sum of the integrated
intensities of these lines.
Small differences between many weak spectral lines in the two models  result in
somewhat different
magnetization curves when $x$ is not very small.

\acknowledgments
This work was supported by CNPQ and FAPESP. Travel funds for Y.~S. were
also provided by FAPESP.

\appendix

\section{\label{a:A}THE CORRECTIVE QUINTETS METHOD}
The CQUIN's method for the \J1 model was described in I.
For a cluster model with two AF exchange constants the CQUIN's method involves
many more steps, and is much more elaborate.
The method is appropriate  when the two exchange constants have
very different magnitudes. This is always the case when  the model is lopsided.

\subsection{\label{a:AA}Corrective and real quintets}
The CQUIN's method is used only in the calculation of $M$, not in the
calculation of \dMdB\ (see Sec.~VI of~\Reftwo). The infinite sum in Eq.~(2) of
\Reftwo\  cannot be carried out.
Therefore, it is truncated after a finite number of terms.
When the CQUIN's method is used, the truncated sum includes only clusters of
sizes $n_c{\leq}5$.
The CQUIN's method replaces the remainder
by an ensemble of fictitious clusters of size $n_c{=} 5$.
These fictitious quintets are the corrective quintets (CQUIN's).

The ensemble of CQUIN's is specified by giving the cluster types of the CQUIN's,
and the cluster population for each type.
The description below is for the \J1-\J2 model, assuming
that the ratio \rJ2 is small.
The CQUIN's method for the \J1-\J3 model is similar.

It is important to distinguish between ``real quintets'' and CQUIN's.
The real quintets are  included in the
finite  truncated sum of Eq.~(2) of \Reftwo\  (hereafter called the ``truncated
sum'').
The CQUIN's are fictitious quintets whose only purpose is to approximate the
remainder.
The cluster types of the CQUIN's are
the same as the cluster types of the real quintets, i.e., they are all the
quintet types of the  \J1-\J2 model.
The remaining task is to specify the population (number of CQUIN's) for each of
these types.

\subsection{\label{a:AB}Guidelines for selecting the populations of CQUIN's}
Two guidelines, or principles, are used to specify the populations of the
various types of CQUIN's.
First, the ensemble of CQUIN's should include a collection of
skeletons which resembles, to the extent possible, the collection of skeletons
in the remainder.
The rationale is that the MST's in the high-field part of the spectrum are
largely controlled by the skeletons.
An inherent limitation is that skeletons of CQUIN's are all of sizes
$n_\Sk{\leq}5$,
whereas some skeletons in the remainder have larger sizes.
The CQUIN's can only approximate the magnetization from the remainder, they cannot
reproduce it exactly.

The second guideline demands that after all the CQUIN's are included, the
calculated magnetization $M$ in the limit $B{\to}\infty$ will have
the true saturation value.
This means that the total number of spins in the  ensemble of CQUIN's
must equal the total number of spins in the remainder.
This number is $P_{>5}N_\mathrm{total}$.

\subsection{\label{a:AC}Role of the parent  \J1 model}
The parent \J1 model plays a major role in the CQUIN's method.
One reason is that the definitions of a simple-skeleton type,  and of a fragment
type, are based on a comparable cluster type  $c$ in  the parent \J1 model.
The comparable  cluster type has the same set of \J1 bonds.
The ``single''
is the only cluster type $c$ of the \J1 model that has no corresponding
skeleton/fragment type \skel{c}.

Consider one realization of any cluster type $c$ of the parent \J1 model, except
a single.
When the \J1 model is replaced by the \J1-\J2 model, the spins that were earlier
in this realization form a simple skeleton, or a fragment, of type \skel{c}.
These skeletons and fragments will be incorporated into many cluster types of
the  \J1-\J2 model.

The skeletons/fragments of type \skel{c} that are incorporated into clusters of
sizes not exceeding $5$ are in the truncated sum.
These skeletons and fragments will be treated exactly.
The other skeletons/fragments of type \skel{c} are incorporated into clusters
larger than quintets, which
are in the remainder. The simple skeletons and fragments of type \skel{c} that
are in the remainder will be represented by those in the CQUIN's.

If possible, the collection of CQUIN's should contain the total number of
skeletons/fragments of type \skel{c}  that are in the remainder.
The limitation is that CQUIN's contain  only simple skeletons of sizes
$n_\Sk{\le}5$, and only fragments of sizes $n_\F{=}2$ or $3$.
Those skeletons and fragments in the remainder which conform to these size
limitations can be well represented  by CQUIN's.
Larger skeletons and fragments in the remainder cannot be well represented by
CQUIN's.

In view of the limitation just mentioned, the only cluster types $c$ of the \J1
model that are considered explicitly
are those with sizes up to $5$.
The population  $N_{c0},$ in the \J1 model,  of each of these cluster  types $c$
is counted at the start of the CQUIN's method.
This count is called the ``initial count'' for type $c$.

Cluster types $c$ of the \J1 model that have sizes $n_c{>}5$  are not treated
explicitly.
However, these cluster types are not totally ignored.
An allowance for \J1-clusters  larger than quintets is made by increasing the
initial counts $N_{c0}$  for all the quintet types $c$ of the \J1 model by
$\Delta N_{c0}$. These $\Delta N_{c0}$  are obtained from
the CQUIN's procedure \emph{for the \J1 model}.\cite{\refone}
The adjustments $\Delta N_{c0}$  are carried out only  for
the quintet types of the \J1 model, not for other cluster types.

\subsection{\label{a:AD} Adding CQUIN's in stages}
The ensemble of CQUIN's that replaces the remainder is assembled in stages.
At each stage CQUIN's with skeletons and fragments of only one type, \skel{c},
are added.
The sequence of stages is discussed later.
The following is a description of any one of these stages.

At the beginning of a stage, all skeletons and fragments of type \skel{c} that
have already been included in the calculation of $M$ are counted.
This count is called  the ``second count'' for type $c$. The result of the
second count will be called $N_c^*$.
In most cases $N_c^*$ includes only the skeletons and fragments of type \skel{c}
that are in the truncated sum.
The only exception is in the second count of skeletons/fragments of type
\skel{2},
performed at the beginning of the stage in which CQUIN's with \skel{2}
skeletons/fragments are added.
As explained below, CQUIN's with \skel3 fragments are added in an earlier stage
which is devoted to \skel3 skeletons and fragments.
What complicates matters is that whenever a \skel3 fragment was added, a \skel2
fragment was automatically added at the same time.
The reason is that a skeleton of a  CQUIN with a \skel3 fragment is always of
type \skel3-\skel2.
The \skel2 fragments that were introduced before the beginning of the stage
devoted to \skel2 skeletons/fragments must be included in the second count for
this stage.

The CQUIN's types that are added at any one stage include all the quintet types
of the \J1-\J2 model that have skeletons and fragments of type \skel{c}.
The populations of these CQUIN's types are determined by two rules:
1) The  total number of skeletons and fragments of type \skel{c}, in all the
CQUIN's that are added at this stage,
is equal  to the difference $(N_{c0}{-}N_c^*)$ between the initial and second
counts for type $c$.
This requirement fulfills the first guideline in part \ref{a:AB} of this
Appendix.
2) The ratios between populations of different CQUIN's types that are added
in any stage must be equal to the population ratios for real quintets of the
same types.

\subsection{\label{a:AE}Order of the various stages of the CQUIN's method}
Each stage of adding  CQUIN's is related to a particular cluster type $c$ of the
\J1 model.
The sequence of stages follows the order of decreasing cluster size $n_c$, in
the \J1 model.
The 10 cluster types of the \J1 model lead to 10 stages.

\subsubsection*{Stages 1--4}
The first four stages treat the four quintet types of the \J1 model:
$5A$, $5B$, $5C$, and $5D$. The order of these four stages is arbitrary.
Consider quintet type $5A$. \Figure2  shows that the second count will only
include the \skel{5A} skeletons in real quintets of types 5-1 up to 5-6.
The CQUIIN's types that are added in this stage are also 5-1 up to  5-6.

Cluster types $5B$, $5C$, and $5D$, of the \J1 model are handled in a similar
way, in three separate stages.
For $5C$, the second count  involves only the \skel{5C} skeletons of real
quintets of type  5-10.
The CQUIN's  that are added  in this stage are also of type 5-10.
A  similar remark applies to the stage devoted to cluster type $5$D of the \J1
model.

\subsubsection*{Stages 5--7}
These stages treat the three quartet types of the \J1 model: $4A$, $4B$, and
$4C$.
Again, the order is arbitrary.
The second count for quartet type $4A$ includes the \skel{4A} skeletons in the
quartet types 4-1, 4-2, and 4-3 (\Fig1), and in real quintets of types 5-12 up
to 5-17 (\Fig2).
The cluster types of the CQUIN's are 5-12 up to 5-17.
Quartet types $4B$ and $4C$ of the \J1 model are handled in a similar way, in
separate stages.

\subsubsection*{Stage 8}
This stage treats the only triplet type (type $3$) of the \J1 model.
The second count includes the \skel3 skeletons of the following  cluster types:
3-1,  3-2,  4-6 up to 4-8, and real quintets of  types 5-21 up to 5-29.
The second count also includes the \skel3 fragments in the \skel3-\skel2
skeletons of real quintets of types 5-36 up to 5-38.
The cluster types of the CQUIN's are 5-21 up to 5-29, and  5-36 up to 5-38.
Note that CQUIN's of  types 5-36 up to 5-38 contain  \skel2 fragments, in
addition to \skel3 fragments.
Therefore, already at this stage some \skel2 fragments are  added
to the ensemble of  CQUIN's.

\subsubsection*{Stage 9}
This stage involves the pairs (cluster type $2$) of the \J1 model.
The second count includes the \skel2 skeletons of the cluster types
2-1,  3-3,  4-9 up to 4-11, and real quintets of types  5-30 up to 5-35.
Also included are the \skel2 fragments of the
\skel2-\skel2 skeletons in quartets of type 4-12, and in real quintets of types
5-39 up to 5-41.
The \skel2 fragments in \skel3-\skel2 skeletons of real quintets of types 5-36
up to 5-38 are also be included in the second count.
Finally, the \skel2 fragments in the CQUIN's of types 5-36  up to 5-38, which
were added in stage No. 8, are  included.
The CQUIN's that are added in stage 9 are of types 5-30 up to 5-35, and 5-39 up
to 5-41.

\subsubsection*{Final Stage: CQUIN's that are pure \J2 clusters}
The final (10th) stage corrects for the ``missing singles'' of the  \J1 model.
The single is the only cluster type of the \J1 model which has no corresponding
skeleton type, or fragment type.
Therefore, the CQUIN's that are added in the final stage have no skeletons; they
are pure \J2 clusters.

The initial count $N_{10}$ is the number of singles in the \J1 model.
When the \J1 model is replaced by the \J1-\J2 model,  the former
singles end up in cluster types of the \J1-\J2 model.
Some remain as singles (cluster type 1-1 in the new \J1-\J2 model).
Some others  end up in pure \J2 clusters, and the rest  end up in spinned
decorations of mixed \J1-\J2 clusters.

The former singles of the \J1 model that end up in clusters of sizes up to 5
are in the truncated sum. These spins are included in the second count $N_1^*$.
The second count also includes the spins in all spinned decorations of the
CQUIN's that were introduced in the earlier stages.
The CQUIN's types that are added in the final (10th) stage are the pure-\J2
quintet
types of the \J1-\J2 model, namely, cluster types 5-42 up to 5-45 in \Fig2.
The total number of spins in these newly added  CQUIN's is $(N_{10}{-}N_1^*)$.

\subsection{\label{a:AG}Example}

\begin{figure}[b]\includegraphics[scale=1]{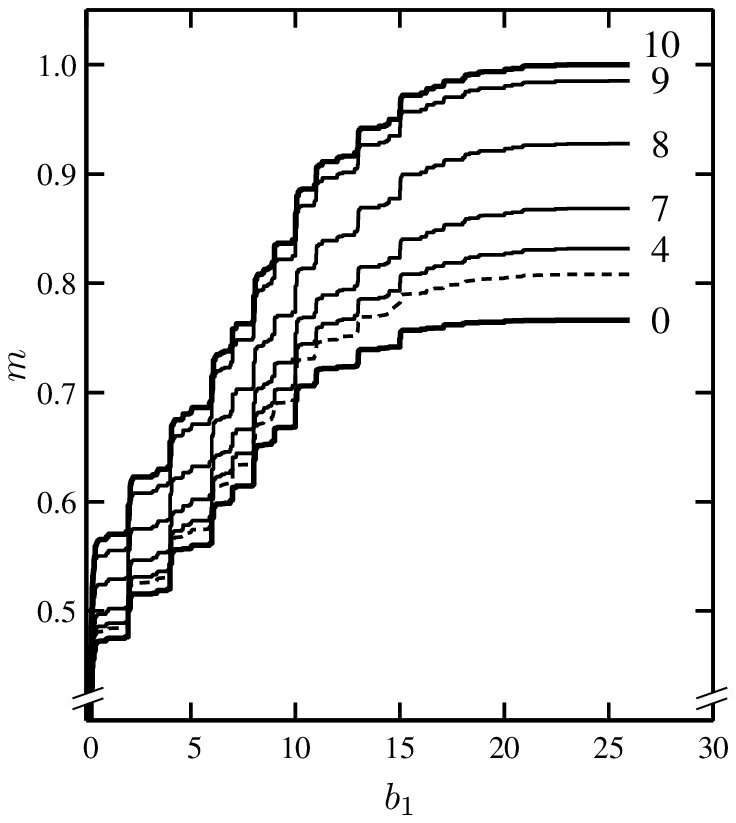}
\caption{\label{Fig13}%
An example of the CQUIN's method, showing the calculated magnetization after the
completion of various stages.
The example is for  the \J1-\J2 model, with $x{=}0.16$, $\irJ2{=}0.028$,  \Spin,
and $T{=}0$.
The abscissa is the primary reduced field $b_1$. The ordinate is the reduced
magnetization $m{=}M{/}M_0$.
The curve labeled as ``0'' represents the truncated sum alone, before  any
CQUIN's are introduced.
The curve labeled as ``10'' is the total magnetization, after all 10 stages of
the CQUIN's method have been completed.
The curves labeled as ``4'',  ``7'', ``8'',  and ``9'', represent the completion
of the 4th, 7th, 8th, and 9th
stages, respectively. The dashed curve is discussed in the text.}
\end{figure}

An example of the application of the CQUIN's method is shown in \Fig{13}.
The various curves represent the reduced magnetization $m$, in the \J1-\J2
model,
after the completion of various stages in the CQUIN's procedure.
The curve labeled as ``0'' represents the truncated sum alone, without any
CQUIN's.
The curve labeled as ``4'' is the calculated magnetization after the completion
of stages 1--4.
The curves labeled as ``7'', ``8'', ``9'', and ``10'', represent the
magnetization after the
completion of stages 5--7, 8, 9, and 10, respectively.
The curve labeled as 10 is the final result.

The dashed curve between the curves ``0'' and ``4''  has the following meaning.
Each of the initial counts $N_{c0}$ for quintets of types $c$ of the \J1 model
contains
two contributions (see section \ref{a:AC} of this appendix).
The first is the  actual initial population count for quintets of this type.
The second is the adjustment $\Delta N_{c0}$, introduced to correct for
neglecting clusters of
the \J1 model that are larger than quintets.

Each of the stages from 1 to 4 corrects for the difference  $(N_{c0}{-}N_c^*)$
between the
initial and second counts for one quintet type $c$.
There are two contributions to this  difference, corresponding to the two
contributions to $N_{c0}$.
The dashed curve in \Fig{13} represents the magnetization that would have been
present at the completion
of stages 1--4 if the only contribution to $N_{c0}$ had been from the adjustment
$\Delta N_{c0}$.

Physically, the difference between the dashed curve and the curve ``0''
approximates
the magnetization from those clusters in the remainder that contain simple
skeletons or fragments
of  sizes larger than 5.
The difference between the curve labeled as ``4'' and the dashed curve
approximates the magnetization
from those clusters in the remainder that  have simple skeletons or fragments
with 5 spins.


\newcommand{\noopsort}[1]{} \newcommand{\printfirst}[2]{#1}
  \newcommand{\singleletter}[1]{#1} \newcommand{\switchargs}[2]{#2#1}

\end{document}